\documentclass[preprint,aps,prd,10pt,nofootinbib,superscriptaddress,preprintnumbers,titlepage,amssymb,amsmath]{revtex4-2}

\usepackage{graphicx}
\usepackage{amssymb,amsfonts,amsmath}
\usepackage{hyperref}
\usepackage{nicefrac}
\usepackage[utf8]{inputenc}
\usepackage[T1]{fontenc}
\usepackage{xcolor}

\graphicspath{{plots/}}

\renewcommand{\l}{\left}
\renewcommand{\r}{\right}
\newcommand{\g}[1]{\gamma_{#1}} 
\newcommand{\DB}[1]{\stackrel{\leftarrow}{D}_{#1}} 
\newcommand{\DF}[1]{\stackrel{\rightarrow}{D}_{#1}} 
\newcommand{\DBF}[1]{\!\stackrel{\leftrightarrow}{D}_{#1}} 
\newcommand{\tr}{\mathrm{tr}}
\newcommand{\bra}[1]{\left< #1 \right|} 
\newcommand{\ket}[1]{\left| #1 \right>} 

\newcommand{\chiral}[1]{\mathring{#1}} 

\newcommand{\gev}{\,\mathrm{GeV}}

\newcommand{\mev}{\,\mathrm{MeV}}
\newcommand{\fm}{\,\mathrm{fm}}

\newcommand{\MSbar}{{\overline{\mathrm{MS}}}}

\newcommand{\phys}{\mathrm{phys}}
\newcommand{\stat}[1]{(#1)_\mathrm{stat}}
\newcommand{\sys}[1]{(#1)_\mathrm{sys}}
\newcommand{\total}[1]{[#1]_\mathrm{total}}
\newcommand{\tex}{t_\mathrm{ex}}
\newcommand{\tins}{t_\mathrm{ins}}

\newcommand{\tsep}{t_\mathrm{sep}}
\newcommand{\tsepmin}{t_\mathrm{sep}^\mathrm{min}}

\newcommand{\tseplo}{t_\mathrm{sep}^\mathrm{lo}}
\newcommand{\tsephi}{t_\mathrm{sep}^\mathrm{hi}}

\newcommand{\avgx}[2]{\langle x \rangle_{#2 u #1 #2 d}}

\newcommand{\vp}{\vec{p}}
\newcommand{\vpi}{\vec{p}_i}
\newcommand{\vq}{\vec{q}}
\newcommand{\vpf}{\vec{p}_f}
\newcommand{\vxi}{\vec{x}_i}
\newcommand{\vxop}{\vec{x}_{op}}
\newcommand{\vxf}{\vec{x}_f}

\newcommand\HIM{\affiliation{Helmholtz~Institute~Mainz, Staudingerweg 18, D-55128 Mainz, Germany}}
\newcommand\GSI{\affiliation{GSI Helmholtzzentrum f\"ur Schwerionenforschung, D-64291 Darmstadt, Germany}}
\newcommand\PRISMA{\affiliation{PRISMA$^+$~Cluster~of~Excellence and Institut~f\"ur~Kernphysik, Johannes~Gutenberg-Universität~Mainz, 55099~Mainz, Germany}}
\begin{document}

 \title{\textbf{Improved analysis of isovector nucleon matrix elements with \texorpdfstring{$N_f=2+1$}{Nf=2+1} flavors of \texorpdfstring{$\mathcal{O}(a)$}{O(a)} improved Wilson fermions}}

 \author{Dalibor Djukanovic} \HIM \GSI
 \author{Georg von Hippel}   \PRISMA
 \author{Harvey~B.~Meyer}    \PRISMA \HIM
 \author{Konstantin~Ottnad}  \email{kottnad@uni-mainz.de}  \PRISMA
 \author{Hartmut~Wittig}     \PRISMA \HIM

 \preprint{MITP-24-014}

 \date{\today}

 \begin{abstract}
  We present an update of our determination of the isovector charges
$g_A^{u-d}$, $g_S^{u-d}$ and $g_T^{u-d}$, and the isovector twist-2
forward matrix elements $\langle x\rangle_{u-d}$, $\langle
x\rangle_{\Delta u-\Delta d}$ and $\langle x\rangle_{\delta u-\delta
d}$ on the $N_\mathrm{f}=2+1$ gauge ensembles generated by the
Coordinated Lattice Simulations (CLS) effort. We have significantly
extended our coverage of the parameter space by adding ensembles at
the physical pion mass and fine lattice spacing, at nearly-physical
pion masses and very fine lattice spacings, and at very large physical
lattice volumes, enabling a well-controlled extrapolation to the
physical point. Another major improvement is achieved owing to the
extended range of source-sink separations, which allows us to perform
two-state fits to summed correlator ratios, leading to a much higher
level of control over excited-state effects. Systematic uncertainties
from the chiral, continuum and infinite-volume extrapolations are
incorporated via model averages based on the Akaike Information
Criterion. Our final results at the physical point are
 $g_A^{u-d} = 1.254\stat{19}\sys{15}\total{24}$,
 $g_S^{u-d} = 1.203\stat{77}\sys{81}\total{112}$,
 $g_T^{u-d} = 0.993\stat{15}\sys{05}\total{16}$,
 $\avgx{-}{}       = 0.153\stat{15}\sys{10}\total{17}$,
 $\avgx{-}{\Delta} = 0.207\stat{15}\sys{06}\total{16}$, and
 $\avgx{-}{\delta} = 0.195\stat{17}\sys{15}\total{23}$.
While our results for the isovector charges are in excellent agreement
with the FLAG\,21 averages, we note that our error for the tensor
charge $g_T^{u-d}$ is considerably smaller.

 \end{abstract}

 \maketitle

\section{Introduction} \label{sec:introduction}

The forward matrix elements of local currents, i.e. the charges, of the nucleon
are some of the most basic structural quantities that can be defined within
QCD. In the case of isovector currents, these charges can be determined from
lattice simulations without the need to consider quark-disconnected diagrams,
which has led to a significant effort within the lattice community to determine
these quantities.
While the isovector axial charge, $g_A^{u-d}=1.2754(13)$
\cite{ParticleDataGroup:2022pth}, is known to high precision
experimentally and serves mostly as a benchmark for lattice QCD calculations,
the isovector scalar and tensor charges $g_S^{u-d}$ and $g_T^{u-d}$ 
are only poorly known from phenomenology,
so that lattice QCD can provide an important first-principles
prediction with significant impact on e.g. dark matter searches
\cite{Cirelli:2013ufw} and searches for Beyond the Standard Model (BSM)
sources of $CP$-violation \cite{Bhattacharya:2015esa}.
Lattice results for these quantities are now being collected by the FLAG
group, which gives global lattice averages \cite{FlavourLatticeAveragingGroupFLAG:2021npn}
for $g_A^{u-d}$ (based on Ref.~\cite{Liang:2018pis,Gupta:2018qil,Chang:2018uxx,Walker-Loud:2019cif,Harris:2019bih}),
$g_S^{u-d}$ \cite{Gupta:2018qil,Harris:2019bih}
and $g_T^{u-d}$ \cite{Gupta:2018qil,Harris:2019bih}.
More recent results (not yet contained in the FLAG\,2021 averages) can be found in
Refs.~\cite{Lin:2020wko,Park:2021ypf,Bali:2023sdi,QCDSFUKQCDCSSM:2023qlx}.

Looking beyond local currents, the forward matrix elements of
twist-2 operators provide access to the average quark momentum
fraction $\langle x\rangle$,
and to the second helicity and transversity moments. For isovector
operators, these can again be determined on the lattice without requiring the
computation of quark-disconnected contributions. Beyond twist-2, lattice
calculations quickly become infeasible due to rapidly declining signal-to-noise
ratios as well as operator mixing.
Lattice results for twist-2 matrix elements have not been included in
the FLAG report until now, and recent
results \cite{Harris:2019bih,Yang:2018nqn,Mondal:2020cmt,Mondal:2020ela}
are fewer than for the charges.

In this paper, we present an update over our previous determination
\cite{Harris:2019bih} of the
isovector axial, scalar and tensor charges of the nucleon, the isovector
average quark momentum fraction, and the second isovector helicity and
transversity moments.
The main improvements over our previous work are: firstly, the addition of several
ensembles, including one at a pion mass slightly below the physical value, one at
an almost physical pion mass at very fine lattice spacing, and two at large
physical volumes; secondly, additional source-sink separations and increased statistics
on almost all ensembles used; thirdly, the use of a two-state fit to the summed ratio,
which combines the explicit treatment of excited-state effects with the
parametric suppression of excited-state contributions inherent to the
summation method \cite{Maiani:1987by,Gusken:1989ad,Bulava:2011yz,Capitani:2012gj},
permitting the use of a much larger fit range while yielding a much better
description of the data; fourth, the implementation of full
$\mathcal{O}(a)$-improvement for the local charges; and finally, 
the use of the Akaike information criterion (AIC) \cite{Akaike1998}
to perform a model average incorporating different
ansätze and cuts for the chiral, continuum and infinite-volume extrapolation.
Taken together, these improvement allow for a significant reduction in both
statistical and systematic errors.

This paper is structured as follows: in section~\ref{section:setup}, we
describe our lattice setup, detailing the observables measured, the ensembles
used, and the computational methods employed. Section~\ref{sec:ESA} details our
excited-state analysis, while section~\ref{sec:CCF} describes the ansätze we use
for the chiral, continuum and infinite-volume extrapolation. Our model
averaging procedure is given in section~\ref{sec:AIC_and_final_results}
together with our final physical results, which are compared to other
determinations in section~\ref{sec:summary} together with some brief conclusions.

\section{Lattice setup} \label{section:setup}
The subject of this study are forward nucleon matrix elements (NMEs) of the form
\begin{equation}
 \bra{N(p_f, s_f)} \mathcal{O}^X_{\mu_1...\mu_n}(x) \ket{N(p_i, s_i)} = e^{i q \cdot x} \bar{u}(p_f,s_f) W^X_{\mu_1...\mu_n}(Q^2) u(p_i, s_i) \,,
 \label{eq:NME} 
\end{equation}
where $N(p_f, s_f)$ ($N(p_i, s_i)$) and $\bar{u}(p_f,s_f)$ ($u(p_i,
s_i)$) denote nucleon states and spinors with initial (final) state
momentum $p_i$ ($p_f$) and spin $s_i$ ($s_f$). The operators
$\mathcal{O}^X_{\mu_1...\mu_n}(x)$ are drawn from the set
$X\in\l\{A,S,T,vD,aD,tD\r\}$ and are defined in Eqs~(\ref{eq:O_A})--(\ref{eq:O_tD})
below. Each choice of the operator insertion $\mathcal{O}^X_{\mu_1...\mu_n}(x)$ results in a distinct form factor decomposition $W^X_{\mu_1...\mu_n}(Q^2)$ on the r.h.s., where $Q_\mu=(i E_f-i E_i, \vq)$, $\vq=\vpf-\vpi$ defines the Euclidean four-momentum transfer. \par

The starting point for the calculation of NMEs in lattice QCD are spin-projected two- and three-point functions 
\begin{align}
 C^\mathrm{2pt}(t_f-t_i, \vp) &= \Gamma_z^{\alpha\beta} \sum_{\vec{x}_f} e^{i \vp \cdot (\vxf-\vxi)} \langle J_{N,\alpha}(\vxf, t_f) \bar{J}_{N,\beta}(\vxi,t_i)\rangle \,, \label{eq:2pt} \\ 
 C^X_{\mu_1...\mu_n}(t_{op}-t_i, t_f-t_i, \vq, \vpf) &= \Gamma_z^{\alpha\beta} \sum_{\vxf, \vxop}  e^{i \vpf \cdot (\vxf-\vxop)} e^{i \vp \cdot (\vxop-\vxi)} \langle J_{N,\alpha}(\vxf, t_f) \mathcal{O}^X_{\mu_1...\mu_n}(\vxop, t_{op}) \bar{J}_{N,\beta}(\vxi, t_i)\rangle \,. \label{eq:3pt}
\end{align}
where $\Gamma_z=\frac{1}{2}(1+\g{0})(1+i\g{5}\g{3})$ and
$J_{N,\alpha}(\vec{x},t)$ denotes a suitable interpolating field for
the nucleon. In this work we restrict ourselves to vanishing
four-momentum transfer $Q^2=0$. Furthermore, we shall assume that
initial and final state are produced at rest, i.e. $\vpf=\vpi=0$, and
drop the corresponding three-momenta from all expressions. Introducing
the usual shorthands $\tsep=t_f-t_i$ and $t=t_{op}-t_i$ for the
pertinent Euclidean time-separations as well as performing an index
shift such that $t_i=0$, the corresponding momentum space expressions for the two- and three-point function in Eqs.~(\ref{eq:2pt})~and~(\ref{eq:3pt}) read
\begin{align}
 C^\mathrm{2pt}(\tsep)         &= \Gamma_z^{\alpha\beta} \langle J_{N,\alpha}(\tsep) \bar{J}_{N,\beta}(0)\rangle \,, \label{eq:2pt_mom_space} \\
 C^X_{\mu_1...\mu_n}(t, \tsep) &= \Gamma_z^{\alpha\beta} \langle J_{N,\alpha}(\tsep) \mathcal{O}^X_{\mu_1...\mu_n}(t) \bar{J}_{N,\beta}(0)\rangle \,. \label{eq:3pt_mom_space} 
\end{align}
The remaining dependence on the source position $t_i$ has been dropped in these expressions as well, assuming that results for multiple source positions are averaged to improve the statistical precision. Note that this requires translational invariance, which has implications for the source placement depending on the choice of boundary conditions, cf. subsection~\ref{subsec:computational_details}. In order to extract matrix elements from three-point functions, unknown overlap factors must be canceled out. At vanishing momentum transfer this is accomplished by forming a simple ratio of the three- and two-point function
\begin{equation}
 R^X_{\mu_1...\mu_n}(t,\tsep) = \frac{C^X_{\mu_1...\mu_n}(t, \tsep)}{C^\mathrm{2pt}(\tsep)} \,.
 \label{eq:ratio}
\end{equation}
Ground-state dominance is achieved at asymptotically large Euclidean time separations, i.e.
\begin{equation}
  \tilde{R}^X_{\mu_1...\mu_n} \equiv \lim_{t\rightarrow\infty} \ \lim_{(\tsep-t)\rightarrow\infty} R^X_{\mu_1...\mu_n}(t,\tsep) = \mathrm{const} \,.
  \label{eq:asymptotic_ratio}
\end{equation}
However, in actual lattice calculations it cannot be guaranteed that
the naive implementation of this limit is free from systematic bias
due to unsuppressed excited-state contributions.
The reason for this is the notorious signal-to-noise problem in nucleon structure calculations which restricts the accessible source-sink separations to values of $\tsep\lesssim 1.5\fm$. Therefore, many different approaches have been developed over the years by various groups attempting to improve the reliability of the ground-state extraction \cite{Ottnad:2020qbw}. The details of our excited-state analysis are discussed in Section~\ref{sec:ESA}. \par

\subsection{Observables} \label{eq:observables}
We consider isovector combinations of the following set of local, dimension-three operator insertions
\begin{align}
 \mathcal{O}^{A}_{\mu}(x)    &= \bar{q}(x) \g{\mu}\g{5} q(x) \,,    \label{eq:O_A} \\
 \mathcal{O}^{S}_{}(x)       &= \bar{q}(x) q(x) \,,                 \label{eq:O_S} \\
 \mathcal{O}^{T}_{\mu\nu}(x) &= \bar{q}(x) \sigma_{\mu\nu} q(x) \,. \label{eq:O_T}
\end{align}
as well as isovector combinations of three twist-2, dimension-four operator insertions, i.e.
\begin{align}
 \mathcal{O}^{vD}_{\mu\nu}(x)     &= \bar{q}(x) \g{\l\{\mu\r.} \DBF{\l.\nu\r\}} q(x) \,,                \label{eq:O_vD} \\
 \mathcal{O}^{aD}_{\mu\nu}(x)     &= \bar{q}(x) \g{\l\{\mu\r.} \g{5} \DBF{\l.\nu\r\}} q(x) \,,          \label{eq:O_aD} \\
 \mathcal{O}^{tD}_{\mu\nu\rho}(x) &= \bar{q}(x) \sigma_{\l[\mu\l\{\nu\r.\r]} \DBF{\l.\rho\r\}} q(x) \,. \label{eq:O_tD}
\end{align}
The second group of operators involves the symmetric derivative $\DBF{\mu}=\frac{1}{2} (\DF{\mu}-\DB{\mu})$ and the notation $\{...\}$ and $[...]$ refers to symmetrization over indices with subtraction of the trace and anti-symmetrization, respectively. Assuming asymptotically large Euclidean time separations as given by Eq.~(\ref{eq:asymptotic_ratio}), the form factor decompositions $W^{A,S,T}_{\mu_1...\mu_n}(Q^2)$ with $n=0,1,2$ for the first group of operators give rise to the local isovector nucleon charges $g_{A,S,T}^{u-d}$ through
\begin{align} 
 \tilde{R}^{A}_{\mu}    &= i\delta_{3\mu} g_A^{u-d} \,,      \label{eq:NME_g_A} \\
 \tilde{R}^{S}          &= g_S^{u-d} \,,                     \label{eq:NME_g_S} \\
 \tilde{R}^{T}_{\mu\nu} &= \epsilon_{03\mu\nu} g_T^{u-d} \,. \label{eq:NME_g_T} 
 \end{align}
Similarly, the isovector, one-derivative operator insertions are related to the isovector average quark momentum fraction $\avgx{-}{}=A^{u-d}_{20}(0)$, helicity momentum $\avgx{-}{\Delta}=\tilde{A}^{u-d}_{20}(0)$ and transversity moment $\avgx{-}{\delta}=A^{u-d}_{T20}(0)$, that are defined from the corresponding, generalized parton distribution functions at vanishing momentum transfer, cf. Ref.~\cite{Hagler:2004yt}. The corresponding decompositions read
\begin{align}
 \tilde{R}^{vD}_{\mu\nu}     &= m \l( \delta_{0\mu} \delta_{0\nu} - \frac{1}{4} \delta_{\mu\nu} \r) \avgx{-}{} \,,                                   \label{eq:NME_vD} \\
 \tilde{R}^{aD}_{\mu\nu}     &= \frac{i m}{2} \l(\delta_{3\mu} \delta_{0\nu} + \delta_{0\mu} \delta_{3\nu}\r) \avgx{-}{\Delta} \,,                   \label{eq:NME_aD} \\
 \tilde{R}^{tD}_{\mu\nu\rho} &= -\frac{i m}{4} \epsilon_{\mu\nu\rho 3} \l(2 \delta_{0\rho} - \delta_{0\nu} - \delta_{0\mu} \r) \avgx{-}{\delta} \,.  \label{eq:NME_tD}
\end{align}
For operator insertions $X=T,vD,aD,tD$ with $n\geq2$ Lorentz indices the data for $R^X_{\mu_1,...\mu_n}(t, \tsep)$ are averaged over all contributing index combinations, resulting in a favorable signal-to-noise ratio compared to the use of just a single index combination. \par

\subsection{Ensembles} \label{subsec:ensembles}

\begin{table}[!t]
 \caption{Gauge ensembles used in this work. Ensembles with open and periodic boundary conditions in time are indicated by superscripts ``$o$'' and ``$p$'', respectively. $M_\pi$ and $M_N$ have been measured on the same set of configurations and the corresponding values of $M_\pi L$ are included as well. $N_\mathrm{conf}$ is the number of gauge configurations measurements, and $N_\mathrm{meas}^\mathrm{max}$ refers to the number of measurements on the largest value of $\tsep$. The range of source-sink separations is given in physical units by $\tseplo$ and $\tsephi$, and $N_{\tsep}$ is the number of source-sink separations on each ensemble, which are increased by a fixed increment (i.e. one or two units of the lattice spacing) between $\tseplo$ and $\tsephi$.}
 \centering
 \setlength{\tabcolsep}{0.5em}
 \begin{tabular}{cccrcccrcccccc}
  \hline\hline
  ID$^\mathrm{BC}$ & $\beta$ & $a/\fm$ & $\frac{T}{a}\times\bigl(\frac{L}{a}\bigr)^3$ & $L/\fm$ & $M_\pi L$ & $M_\pi/\mev$ & $M_N/\mev$ & $N_\mathrm{conf}$ & $N_\mathrm{meas}^\mathrm{max}$ & $\tseplo/\fm$ & $\tsephi/\fm$ & $N_{\tsep}$ \\
  \hline\hline                                                                                                          
  H102$^o$ & 3.40 & 0.0855 &  $96 \times 32^3$ & 2.74 & 4.99 & 360(3) & 1116(10) & 2037 &  32592 & 0.35 & 1.47 & 14 \\  
  H105$^o$ &      &        &  $96 \times 32^3$ & 2.74 & 3.92 & 283(4) & 1030(14) & 1027 &  49296 &      &      & 14 \\  
  N101$^o$ &      &        & $128 \times 48^3$ & 4.11 & 5.89 & 283(2) & 1038(10) & 1593 &  50976 &      &      & 14 \\  
  C101$^o$ &      &        &  $96 \times 48^3$ & 4.11 & 4.74 & 228(2) &  989(08) & 2000 &  64000 &      &      & 14 \\  
  \hline                                                                        
  S400$^o$ & 3.46 & 0.0756 & $128 \times 32^3$ & 2.42 & 4.33 & 353(3) & 1132(09) & 2873 &  45968 & 0.31 & 1.53 &  9 \\  
  N451$^p$ &      &        & $128 \times 48^3$ & 3.63 & 5.31 & 289(2) & 1054(07) & 1011 & 129408 &      &      &  9 \\  
  D450$^p$ &      &        & $128 \times 64^3$ & 4.84 & 5.35 & 218(2) &  981(09) &  500 &  64000 &      &      & 17 \\  
  \hline                                                                        
  N203$^o$ & 3.55 & 0.0636 & $128 \times 48^3$ & 3.06 & 5.41 & 349(3) & 1118(09) & 1543 &  24688 & 0.26 & 1.41 & 10 \\  
  S201$^o$ &      &        & $128 \times 32^3$ & 2.04 & 3.05 & 295(3) & 1134(10) & 2092 &  66944 &      &      & 10 \\  
  N200$^o$ &      &        & $128 \times 48^3$ & 3.06 & 4.36 & 282(2) & 1061(14) & 1711 &  20532 &      &      & 10 \\  
  D200$^o$ &      &        & $128 \times 64^3$ & 4.07 & 4.27 & 207(2) &  976(09) & 1999 &  63968 &      &      & 10 \\  
  E250$^p$ &      &        & $192 \times 96^3$ & 6.11 & 4.03 & 130(1) &  942(07) &  399 & 102144 &      &      & 10 \\  
  \hline                                                                        
  N302$^o$ & 3.70 & 0.0493 & $128 \times 48^3$ & 2.37 & 4.20 & 350(3) & 1146(12) & 2201 &  35216 & 0.20 & 1.40 & 13 \\  
  J303$^o$ &      &        & $192 \times 64^3$ & 3.16 & 4.24 & 265(2) & 1043(08) & 1073 &  17168 &      &      & 13 \\  
  E300$^o$ &      &        & $192 \times 96^3$ & 4.74 & 4.22 & 176(1) &  971(09) &  569 &  18208 &      &      & 13 \\  
  \hline\hline
 \end{tabular}
 \label{tab:ensembles}
\end{table}

Our lattice calculations are performed on a set of 15 gauge ensembles
listed in Table~\ref{tab:ensembles}. These ensembles have been
generated by the Coordinated Lattice Simulation (CLS) consortium
\cite{Bruno:2014jqa} with $N_f=2+1$ flavors of non-perturbatively
$\mathcal{O}(a)$-improved Wilson fermions \cite{Sheikholeslami:1985ij}
and the tree-level Symanzik-improved gauge action
\cite{Luscher:1984xn}. Since the simulations have been carried out
with a twisted mass regulator in the light quark sector to suppress
exceptional configurations \cite{Luscher:2012av} and the rational
approximation \cite{Clark:2006fx} for the strange quark, the
computation of physical observables requires reweighting. For all but
one ensemble (E300) we make use of the reweighting factors that have
been computed using exact low mode deflation in
Ref.~\cite{Kuberski:2023zky}. The reweighting factors for E300 have
been determined by the conventional method based on a stochastic
estimator as discussed in Ref.~\cite{Bruno:2014jqa}. Furthermore, we
employ the procedure introduced in Ref.~\cite{Mohler:2020txx} to deal
with violations of the positivity of the fermion determinant that
occurs on a small subset of gauge configurations on some of our
ensembles. The majority of ensembles in Table~\ref{tab:ensembles} has
been generated with open boundary conditions (oBC) in the time direction
to prevent topological freezing \cite{Luscher:2011kk,Luscher:2012av},
however, three ensembles (E250, D450 and N451) feature periodic
boundary conditions (pBC). Moreover, all ensembles in
Table~\ref{tab:ensembles} lie on a single chiral trajectory subject to
the constraint $\tr[M] = 2m_l+m_s =\mathrm{const}$, where $M$ denotes
the bare quark mass matrix. \par

While a subset of these ensembles had already been analyzed in a previous study in Ref.~\cite{Harris:2019bih}, there are several important advances, including but not limited to
\begin{enumerate}
 \item The addition of two fine and large boxes in the vicinity of physical quark mass (E250 and E300), as well as two ensembles with $M_\pi L > 5$ and large physical volume at $M_\pi\approx220 \mev$ (D450) and $M_\pi\approx 280\mev$ (N101). These newly added ensembles improve our control over the physical extrapolation, particularly for the chiral extrapolation and finite volume effects.
 \item An increased number of source-sink separations, including values $\tsep<1\fm$ and filling in odd values of $\tsep/a$ for all ensembles at the coarsest lattice spacing ($\beta=3.40$) as well as on D450. This enables a much more fine-grained control of the excited-state contamination.
 \item Increased statistics on various ensembles (e.g. roughly doubled
   gauge statistics on D200, J303, S400) and replacing the N401
   ensemble (which had open boundary conditions) with the newly generated
   N451 ensemble which features periodic boundary conditions and an
   order of magnitude higher statistics.
\end{enumerate}
In particular the inclusion of ensembles down to physical quark masses necessitates also a change of our analysis strategy for the treatment of excited states that is discussed in detail in Section~\ref{sec:ESA}. \par


The dimensionful quantities that enter our analysis are expressed in units of the gradient flow scale $t_0$ \cite{Luscher:2010iy}. To this end we employ the values for $t_0^\mathrm{sym}/a^2$ at the symmetrical point as given in Table~III in Ref.~\cite{Bruno:2016plf}. In order to set the scale in our simulations, we use the world average estimate given by FLAG in Ref.~\cite{FlavourLatticeAveragingGroupFLAG:2021npn}
\begin{equation}
 \sqrt{t_0^\phys}=0.14464(87)\fm \,,
 \label{eq:sqrt_t0phys}
\end{equation}
for the physical value of $t_0$ with $N_f=2+1$ dynamical quark flavors. However, the scale setting affects the final, physical results only through the definition of the physical point in the (light) quark mass, cf. Section~\ref{sec:CCF}, because no explicit conversion to physical units is required for the NMEs. Furthermore, due to this choice of the scale setting procedure the values for $a$ in units of $\fm$ in Table~\ref{tab:ensembles} do not actually enter the analysis. They have been computed using the value in Eq.~(\ref{eq:sqrt_t0phys}) together with the values for $t^\mathrm{sym}/a^2$ to give an indication for the lattice spacing at each value of $\beta$. The value in Eq.~(\ref{eq:sqrt_t0phys}) has also been used to convert the dimensionful quantities in Table~\ref{tab:ensembles} to physical units. \par

\subsection{Computational details} \label{subsec:computational_details}
The computational setup for the quark-connected two- and three-point functions in Eqs.~(\ref{eq:2pt})~and~(\ref{eq:3pt}) is similar to what we have used in various other studies of nucleon structure published in Refs.~\cite{Djukanovic:2021cgp,Agadjanov:2023efe,Djukanovic:2023beb}. The calculations are carried out on point sources for a common choice of the nucleon interpolating operator
\begin{equation}
 J_{N,\alpha}(\vec{x}, t) = \epsilon_{abc} \l(\tilde{u}_a^T(x) C \g{5} \tilde{d}_b(x)\r) \tilde{u}_{c,\alpha}(x) \,
 \label{eq:interpolating_operator}
\end{equation}
where the tilde on the quark fields denotes that Gaussian smearing \cite{Gusken:1989ad} with spatially APE-smeared gauge links \cite{APE:1987ehd} has been applied. At every value of $\beta$, the parameters are tuned such that the resulting smearing radius takes a value of $\sim 0.5\fm$ \cite{vonHippel:2013yfa}. \par

The two- and three-point functions are evaluated on a common set of point sources leading to a statistically favorable signal for the ratio in Eq.~(\ref{eq:ratio}). Besides, this allows us to reuse the forward propagators from the two-point function computation for the corresponding three-point functions at multiple values of $\tsep$, as the three-point functions are computed by the sequential inversion through the sink. Furthermore, the computational cost is reduced by a factor of $\sim 2$ to $\sim 5$ due to the use of the truncated solver method \cite{Bali:2009hu,Blum:2012uh,Shintani:2014vja} for the required inversions of the Dirac operator. The source setup depends on the type of boundary conditions in time for any given ensemble in Table~\ref{tab:ensembles}. On ensembles with pBC the sources can be randomly distributed for every configuration over the entire volume of the lattice, subject only to the constraint resulting from the combination of the truncated solver method with the Schwartz alternating procedure (SAP) preconditioning \cite{Luscher:2003vf,vonHippel:2016wid}. The value $N_\mathrm{meas}^\mathrm{max}$ in Table~\ref{tab:ensembles} refers to the number of measurements on the largest value of $\tsep$. For decreasing values of $\tsep$ the number of measurements, $N_\mathrm{meas}$, is divided by two every one or two step(s) in $\tsep/a$. Due to this choice of downscaling the number of measurements, the resulting signal-to-noise behavior is much less $\tsep$-dependent compared to the unmitigated exponential decay expected when keeping the number of measurements constant as a function of $\tsep$. This prevents giving undue statistical weight to data at small values of $\tsep$ in fits, while the computational cost is significantly decreased at smaller values of $\tsep$ as a side effect. On the other hand, for ensembles with oBC the sources are always located at a single time slice in the bulk of the lattice. In this case the scaling of the number of measurements is only applied for $\tsep\lesssim 1\fm$. Furthermore, the three-point function measurements at $\tsep\gtrsim 1\fm$ have been generated on a fixed set of source positions on most of the ensembles that had been included in the study in Ref.~\cite{Harris:2019bih}. However, for the newer data on E300, N101 and H102, the spatial coordinates of the sources have been distributed randomly on every gauge field. The latter also holds for the spatial components of the sources used for the measurements at $\tsep\lesssim 1\fm$ on all ensembles with oBC. \par

The analysis on individual ensembles is carried out using the jackknife method with pre-binning to account for autocorrelations in the data. However, the NME data itself is essentially unaffected by autocorrelations, it is only the errors for $M_\pi$ and $M_N$ that exhibit relevant effects of autocorrelations on some ensembles. Still, the contributions from $M_\pi$ and $M_N$ to the total error of the NMEs at the physical point is almost negligible. In particular $M_N$ only enters the twist-2 NMEs as a linear coefficient in their respective form factor decompositions in Eqs.~(\ref{eq:NME_vD})-(\ref{eq:NME_tD}), but with a statistical error that is typically an order of magnitude smaller than the error on the NME data itself. After extracting the results for the ground-state NMEs on individual ensembles, it is necessary to combine them in global fits for the physical extrapolations, cf. Section~\ref{sec:CCF}. To this end we employ a parametric bootstrap to include the data from different ensembles in these fits while preserving the correlations between e.g. $M_\pi$ and the NMEs on a given ensemble. Furthermore, the errors on $t_0^\mathrm{sym}/a^2$ from Ref.~\cite{Bruno:2016plf}, the error on the scale itself in Eq.~(\ref{eq:sqrt_t0phys}) and the errors on the renormalization factors are propagated to the final, physical results through the parametric bootstrap procedure. All global fits are then carried out on $N_B=10000$ bootstrap samples to estimate the statistical errors on the physical results. The final errors for the physical NMEs including systematic effects are obtained from model averaging in the last analysis step described in Section~\ref{sec:AIC_and_final_results}. \par

As discussed previously in the context of isoscalar observables in Refs.~\cite{Agadjanov:2023efe,Djukanovic:2023beb,Djukanovic:2023jag}, we observe that measurements on a few point sources on a very small number of gauge configurations stand out as extreme outliers with respect to the distribution across configurations. Including these measurements would lead to unreasonably inflated statistical errors and spoil the scaling with respect to the number of measurements and the value of $\tsep$ for an affected observable on a given ensemble. However, this only occurs on a handful of ensembles, and the issue is generally much less prominent than what has been observed for isoscalar NMEs. Nevertheless, we still employ a similar procedure as that described in the supplemental material of Ref.~\cite{Agadjanov:2023efe}: First, before carrying out the actual analysis, we generate single-elimination jackknife samples for the effective form factors of each NME on every ensemble. In a second step, we scan these samples for ``outliers'' that are more than $\sim6\sigma$ away from the center of the distribution. Subsequently, all configurations that have been flagged in any observable for any value of $\tsep$ and $\tins$ are removed from the final analysis. In total we find seven configurations on five affected ensemble (i.e. a single configuration on E250, D200, N200 and S201, and three configurations of N101), which is reflected by $N_\mathrm{conf}$ and $N_\mathrm{meas}^\mathrm{max}$ in Table~\ref{tab:ensembles}. \par

\subsection{Renormalization} \label{subsec:renormalization}
Apart from $g_A^{u-d}$, all NMEs considered in this study generally require renormalization. However, as the fermion discretization used in the generation of the CLS gauge ensembles breaks chiral symmetry, $g_A^{u-d}$ requires renormalization at finite values of $a$ as well. To this end we make use of the values for $Z_A$ that have been determined for all four values of $\beta$ in Ref.~\cite{DallaBrida:2018tpn} from the chirally rotated Schr\"odinger functional, whereas for the five other operator insertions in Eqs.~(\ref{eq:O_S})-(\ref{eq:O_tD}) we use the values for $Z_S^\MSbar$, $Z_T^\MSbar$, $Z_{v2b}^\MSbar$, $Z_{r2a}^\MSbar$, and $Z_{h1a}^\MSbar$ that have been computed in the $\MSbar$ scheme at a scale of $\mu=2\gev$ in Ref.~\cite{Harris:2019bih}. We remark, that for each of the twist-2 operators only one out of two possible irreps contributes. An important difference to Ref.~\cite{Harris:2019bih} is that full $\mathcal{O}(a)$-improvement has now become available for all three local operator insertions. While there is no contribution from current improvement at zero momentum transfer for this matrix elements, the renormalization pattern changes to 
\begin{equation}
 Z_{X}^\mathrm{imp}(g_0^2, m_q, \bar{m}) = Z_{X}(g_0^2)\l(1 + am_q b_{X}(g_0^2) + 3 a \bar{m}\tilde{b}_{X}(g_0^2) \r) \,, \quad X=A,S,T \,,
 \label{eq:_Z_X_imp}
\end{equation}
where
$m_q=\frac{1}{2a}\bigl(\frac{1}{\kappa_q}-\frac{1}{\kappa_\mathrm{crit}}\bigr)$ denotes the bare subtracted quark mass for $q=l,s$ and $\bar{m}=\frac{1}{3}\l(2m_l+m_s\r)$ is the (bare) average quark mass. The values of $\kappa_\mathrm{crit}$ have been given in Ref.~\cite{Gerardin:2018kpy}, and the improvement coefficients $b_{X}$ and $\tilde{b}_{X}$ for all three local operators have been published by the Regensburg group in their recent study of octet baryon isovector charges in Ref.~\cite{Bali:2023sdi}. In our previous study in Ref.\cite{Harris:2019bih} only partial improvement was implemented for the axial vector matrix element, i.e. neglecting $\tilde{b}_A$ and using an older set of values for $b_A$ from Ref.~\cite{Korcyl:2016ugy}, whereas for the scalar and tensor matrix elements none of the improvement coefficients were available at the time. \par

\section{Excited-state analysis} \label{sec:ESA}
\begin{figure}[!htb]
 \centering
  \includegraphics[totalheight=0.225\textheight]{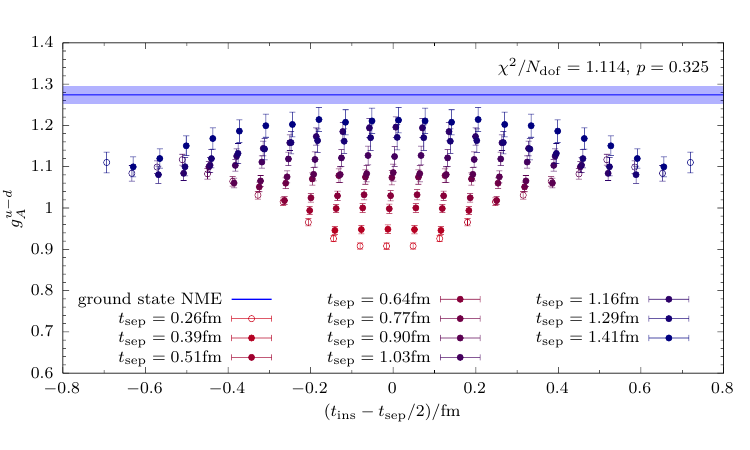}
  \includegraphics[totalheight=0.225\textheight]{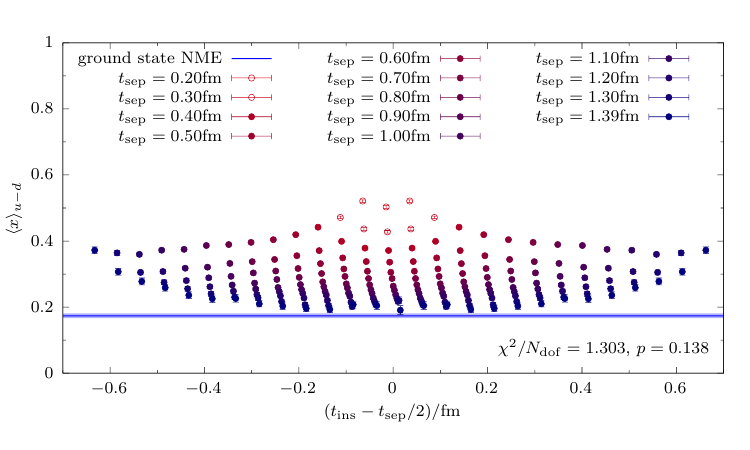} \\
  \includegraphics[totalheight=0.225\textheight]{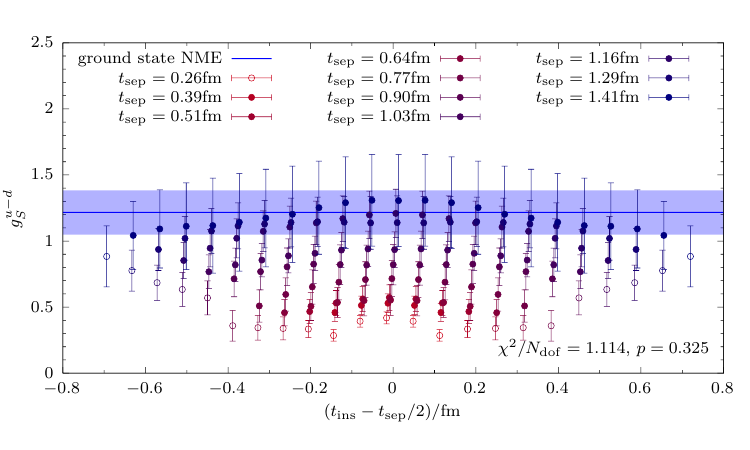}
  \includegraphics[totalheight=0.225\textheight]{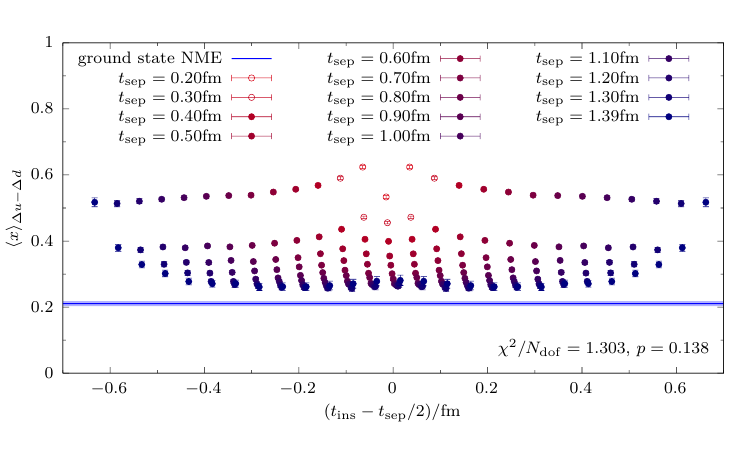} \\
  \includegraphics[totalheight=0.225\textheight]{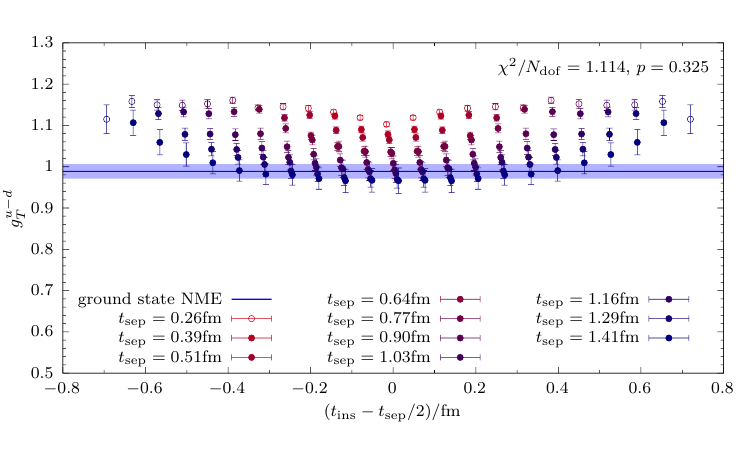}
  \includegraphics[totalheight=0.225\textheight]{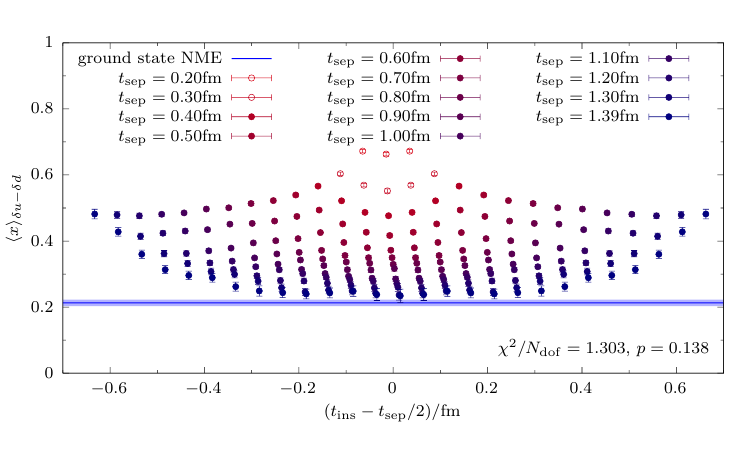} \\
  \caption{Example data for effective form factors of the six observables on the two most chiral ensembles including all available values of $\tsep$. Data for local (twist-2) operator insertions are shown for E250 (E300), respectively. Results with statistical errors for the respective ground state NMEs are indicated by the solid blue line and band. They are obtained from the two-state truncated summation method fit ansatz in Eq.~(\ref{eq:summation_two_state}) for a choice of $\tsepmin\approx0.4\fm$. Data with open symbols do not contribute to the sums that enter the fit due to the choice of $\tsepmin$ or the constraint $\tex/a=1$. The fits to the summed plateau data are carried out simultaneously on any given ensemble with a common parameter for the energy gap $\Delta$ within each of the two sets of local and twist-2 NMEs.}
  \label{fig:eff_FF}
\end{figure}
The excited-state suppression in the ratio method defined by Eqs.~(\ref{eq:ratio})~and~(\ref{eq:asymptotic_ratio}) is insufficient at the accessible values of $\tsep$. This is reflected by the $\tins$ and $\tsep$-dependence of the effective form factors that is displayed in Fig.~(\ref{fig:eff_FF}) for the six isovector NMEs on the two most chiral ensembles. In order to improve the suppression of excited states and reduce the residual contamination we make use of a variant of the summation method \cite{Maiani:1987by,Dong:1997xr,Capitani:2012gj} that is based on the sum of the ratio in Eq.~(\ref{eq:ratio}) over insertion times
\begin{equation}
  S^X_{\mu_1...\mu_n}(\tsep,\tex)\equiv\sum_{\tins=\tex}^{\tsep-\tex} R^X_{\mu_1...\mu_n}(\tins, \tsep) \,.
  \label{eq:summed_ratio}
\end{equation}
Plugging the expressions for the two-state truncation for the nucleon two- and three-point functions
\begin{align}
 C^\mathrm{2pt}(\tsep) =& \l|A_0\r|^2 e^{-m_0 \tsep} + \l|A_1\r|^2 e^{-m_1 \tsep} + ... \,, \label{eq:2pt_two_state_truncation} \\
 C^X_{\mu_1...\mu_n}(\tins, \tsep) \notag =& |A_0|^2 M_{00} e^{-m_0\tsep} + A_0 A_1^* M_{01} e^{-m_0(\tsep-\tins)} e^{-m_1\tins} \notag \\
  &+ A_1 A_0^* M_{10} e^{-m_1(\tsep-\tins)} e^{-m_0\tins} + |A_1|^2 M_{11} e^{-m_1\tsep} + ... \,. \label{eq:3pt_two_state_truncation}
\end{align}
into Eq.~(\ref{eq:ratio}), the corresponding expression for the ratio reads
\begin{equation}
 R^X_{\mu_1...\mu_n}(\tins, \tsep) = \frac{M_{00} + M_{01} \frac{A_1^*}{A_0^*} e^{-\Delta\tins} + M_{01} \frac{A_1}{A_0} e^{-\Delta(\tsep-\tins)} + M_{11} \frac{|A_1|^2}{|A_0|^2} e^{-\Delta\tsep}}{1 + \frac{|A_1|^2}{|A_0|^2} e^{-\Delta\tsep}} \,,
 \label{eq:ratio_two_state_truncation}
\end{equation}
where $\Delta=m_1-m_0$ denotes the energy gap between the ground state and first excited state, and we have exploited the fact that $M_{01}=M_{10}$. The two-state truncation of the summed ratio in Eq.~(\ref{eq:summed_ratio}) is obtained from an expansion for small values of $e^{-\Delta\tsep}$
\begin{align}
 S^X_{\mu_1...\mu_n}(\tsep,\tex) = & M_{00}(\tsep - 2\tex + a) + 2 \tilde{M}_{01} \frac{ e^{-\Delta\tex} - \bigl( e^{\Delta(\tex-a)} + \frac{|A_1|^2}{|A_0|^2} e^{-\Delta\tex}\bigr) e^{-\Delta\tsep}}{1-e^{-\Delta a}} \notag \\
                                   & +\tilde{M}_{11} e^{-\Delta\tsep} (\tsep - 2\tex + a) + \mathcal{O}(e^{-2\Delta\tsep}) \,.
 \label{eq:summation_two_state_truncation}
\end{align}
where we have defined $\tilde{M}_{01}=2\mathrm{Re}\l[ A_1/A_0\r] M_{01}$ and $\tilde{M}_{11}=|A_1|^2/|A_0|^2 (M_{11}-M_{00})$. Neglecting all terms $\sim e^{-\Delta\tsep}$ on the r.h.s. the summation method is recovered in its standard form without explicitly parameterizing the contribution of the first excited state. \par

At our current level of precision we find that terms $\sim \frac{|A_1|^2}{|A_0|^2}$ are not constrained by the data. Therefore, we neglect these contributions in our final fit model
\begin{equation}
 S(\tsep, \tex=a) = M_{00}(\tsep - a) + 2 \tilde{M}_{01} \frac{ e^{-\Delta a} - e^{-\Delta\tsep} }{1-e^{-\Delta a}} \,.
 \label{eq:summation_two_state}
\end{equation}
where we choose $\tex/a=1$ to avoid contact terms. In principle this expression could be fitted simultaneously in all six observables with $\Delta$ as a common fit parameter. However, we observe that allowing for a different gap for the local and the twist-2 matrix elements greatly increases the fit quality, particularly when including smaller values of $\tsep$ in the fit. Therefore, we decided to fit the local and the twist-2 NMEs separately. Still, exploiting the correlations between observables and effectively reducing the number of (nonlinear) fit parameters improves the stability and achievable precision of the results. \par

We remark that the summation-based approach has several features that make it more appropriate for our current set of data than e.g. the two-state ratio fit model that has been used in Ref.~\cite{Harris:2019bih}
\begin{equation}
 R(\tins, \tsep) = c_0 + c_1 (e^{-\Delta\tins} - e^{-\Delta(\tsep-\tins)}) + c_2 e^{-\Delta\tsep} \,,
 \label{eq:ratio_fit_two_state}
\end{equation}
where $c_0=M_{00}$ and data are fitted as a function of $\tsep\geq\tsepmin$ and $\tins\in\left[\tsepmin/2,\tsep/2\right]$. First of all, the leading correction in this fit model behaves as $\mathcal{O}(e^{-\Delta \tsepmin/2})$, whereas for the summation method it is $\mathcal{O}(e^{-\Delta \tsepmin})$. The enhanced excited state suppression is an important advantage of the summation method as it allows to include data at smaller values of $\tsep$ that are more precise and / or numerically cheaper to compute. However, the weaker suppression of excited states by the model in Eq.~(\ref{eq:ratio_fit_two_state}) becomes a real issue for ensembles with $M_\pi\lesssim 200\mev$. Imposing the same criterion on the choice of $\tsepmin$ for the fit ranges of $\tsep$ and $\tins$ that has been used Ref.~\cite{Harris:2019bih} would eliminate almost all data on our two most chiral ensembles, effectively preventing any meaningful fit. Besides, fit models based on the summation method inherently involve (far) fewer degrees of freedom, resulting in smaller covariance matrices, which can improve the stability of the fits. \par

\begin{figure}[!htb]
 \centering
  \includegraphics[totalheight=0.226\textheight]{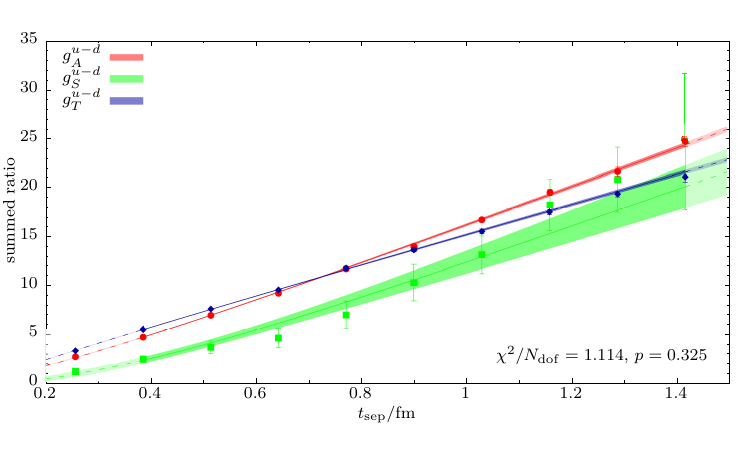}
  \includegraphics[totalheight=0.226\textheight]{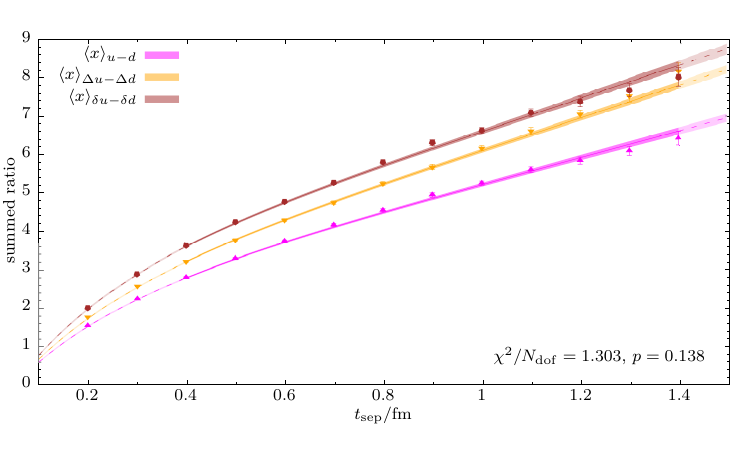} \\
  \includegraphics[totalheight=0.226\textheight]{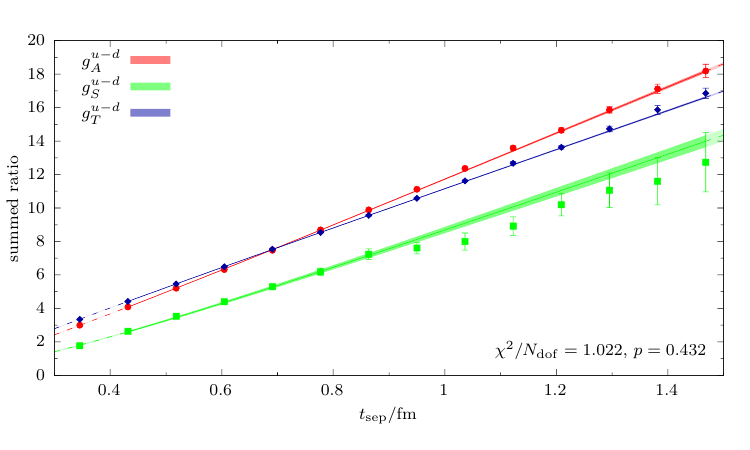}
  \includegraphics[totalheight=0.226\textheight]{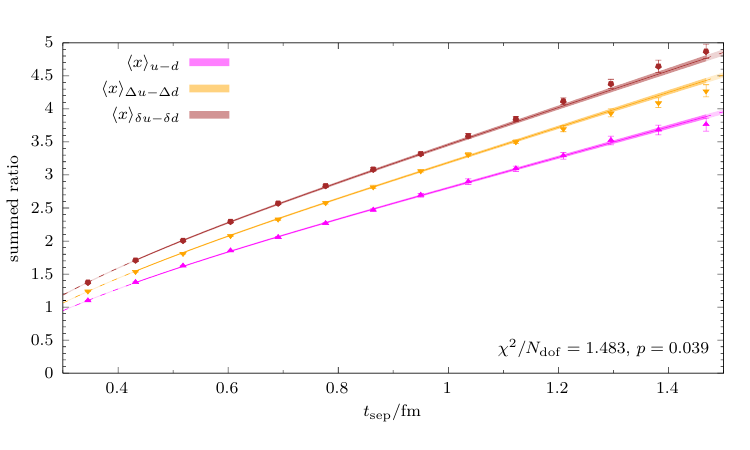} \\
  \caption{Examples of simultaneous fits of the two-state truncated
    summation method fit model in
    Eq.~(\ref{eq:summation_two_state}). Top row: Simultaneous fits of the isovector nucleon charges on the E250 ensemble (left panel)
    and simultaneous fits of twist-2 NMEs on E300 (right panel) corresponding to the data shown in the two columns in
    Fig.~\ref{fig:eff_FF}. Bottom row: local and twist-2 NMEs on the C101 ensemble. The respective fit ranges in $\tsep$ are indicated
    by the solid lines and dark shaded parts of the error bands, whereas the dashed lines and light-shaded error bands represent an
    extrapolation without actual support of the lattice data.}
  \label{fig:two_state_stummation_fits}
\end{figure}

The fits are implemented using the VARPRO method \cite{doi:10.1137/0710036} that we find to greatly improve their robustness as it removes any potential dependence on initial values for the linear fit parameters. The only nonlinear fit parameter is the energy gap $\Delta$, which is generally treated as a free parameter in these fits. In practice, $\Delta$ acts as an effective fit parameter collecting residual contributions from higher states as well. Its statistical precision rapidly deteriorates for increasing value of $\tsepmin$, while its value becomes compatible with zero within large errors at around $\tsepmin \gtrsim 0.8\fm$ even for the ensembles with the statistically most precise data. We remark that for the fits of the local charges on E250 a prior for $\Delta$ is required to stabilize the fit for certain choices of $\tsepmin$. To this end, we employ as a prior the value for $\Delta$ obtained from a simultaneous fit to all six NMEs with $\tsepmin=8a$ (i.e. before the signal is lost in noise) with a $20\%$ width. We have checked that the results are independent of the specific value of the prior within reasonable variations; the prior itself is needed merely to prevent the fit from drifting to obviously unphysical results in some cases. The simultaneous fits to the twist-2 NMEs on E250 exhibit stable convergence without a prior and no further priors are required anywhere else in the analysis. \par

Fig.~\ref{fig:two_state_stummation_fits} shows examples of fits to our lattice data based on Eq.~(\ref{eq:summed_ratio}). The band in each panel represents the result of a simultaneous fit to the summed ratio data for the three NMEs shown in the plot. A key feature observed on all our ensembles is that the deviation from the linear behavior in $\tsep$ becomes highly significant at small values of $\tsep$ due to the excited state contamination. Nevertheless, the fit model in Eq.~(\ref{eq:summation_two_state}) is sufficient to describe the curvature in the data even for choices of $\tsepmin \ll 0.8\fm$. In fact, while the fits in Fig.~\ref{fig:two_state_stummation_fits} have been carried out for $\tsepmin \approx 0.4\fm$, the extrapolation of the fit band typically describes the data very well even at $\tsep<\tsepmin$, down to the smallest available values of $\tsep$, as can be seen in e.g. the top right panel for the twist-2 NMEs on E300. Generally, the resulting curvature of the fit band is strongly dependent on the matrix element: For $g_A^{u-d}$ and $g_S^{u-d}$ it is opposite to $g_T^{u-d}$ as well as the twist-2 NMEs, which all three exhibit a very similar $\tsep$-dependence. The latter is also reflected by rather large correlations between data for different twist-2 operators insertions on the same ensemble. The curvature in the summed ratio data coincides with the behavior of the effective form factors data in Fig.~\ref{fig:eff_FF}, i.e. the effective form factors of $g_A^{u-d}$ and $g_S^{u-d}$ increase as a function of $\tsep$, whereas for the other NMEs they show a monotonic decrease. \par

\begin{figure}[!htb]
 \centering
  \includegraphics[totalheight=0.226\textheight]{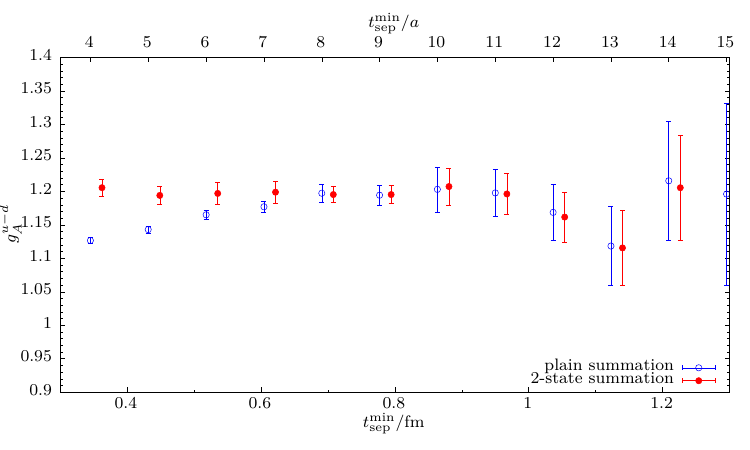}
  \includegraphics[totalheight=0.226\textheight]{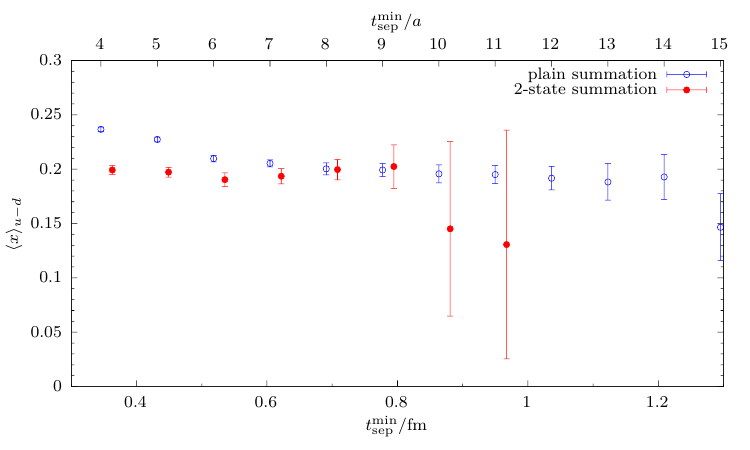}
  \caption{Comparison of the $\tsepmin$ dependence for the plain summation method (open, blue symbols) and two-state summation method (filled, red symbols) fit models on the C101 ensemble. Left panel: $g_A^{u-d}$, right panel: $\avgx{-}{}$. The filled symbols at $\tsepmin\approx 0.4\fm$ in both figures are obtained from the corresponding fits shown in the lower two panels of Fig.~\ref{fig:two_state_stummation_fits}. Data for the two-state fit model are displaced horizontally for clarity.}
  \label{fig:t_sep_min_comparison}
\end{figure}

The lower two panels of Fig.~\ref{fig:two_state_stummation_fits} show similar fits for the six NMEs on the C101 ensemble, which exhibits statistically very precise data and a large number of $\tsep$ values while still being reasonably chiral. Therefore, this ensemble is well suited to demonstrate the efficacy of the two-state fit ansatz. In Fig.~\ref{fig:t_sep_min_comparison} we compare results for $g_A^{u-d}$ and $\avgx{-}{}$ from the two-state ansatz in Eq.~(\ref{eq:summation_two_state}) with the plain, linear summation method as a function of $\tsepmin$. Clearly, the two-state fit ansatz appears to have converged already at the lowest available $\tsepmin$ values in agreement with the results from the plain summation method at around $\tsepmin\gtrsim 0.7\fm$. However, there can be some fluctuations for different values of $\tsepmin$, as can be seen in the right panel for $\avgx{-}{}$ at e.g. $\tsepmin\approx 0.5\fm$. In order to account for any potential impact of such fluctuations on the final results, we employ three sets of data based on fits corresponding to $\tsepmin\in\l\{0.2\fm, 0.3\fm, 0.4\fm\r\}$ across all ensembles. For each of these three data sets we individually carry out the various physical extrapolations described in the next section, before combining all results for any given observable in a model average, cf. Sec.~\ref{sec:AIC_and_final_results}. \par

\section{Physical extrapolation} \label{sec:CCF}

The physical extrapolation of the ground-state NMEs is carried out using fit models that are based on the NNLO expression for the axial charge in $SU(2)$ baryon chiral perturbation theory ($\chi$PT) \cite{Kambor:1998pi}. Our most general fit ansatz for the chiral, continuum and finite volume (CCF) extrapolation reads 
\begin{equation}
 O(M_\pi, a, L) = A_O + B_O M_\pi^2 + \frac{A_O \delta_O}{\l(2 \pi f_\pi\r)^2} M_\pi^2 \log{M_\pi} + C_O M_\pi^3 + D_O a^{n(O)} + E_O \frac{M_\pi^2}{\sqrt{M_\pi L}} e^{-M_\pi L} \,,
 \label{eq:NNLO_CCF_model} 
\end{equation}
where $A_O$, $B_O$, $C_O$, $D_O$ and $E_O$ are treated as free, observable-dependent parameters of the fit. For the pion decay constant we use $f_\pi=130.2(1.2)\mev$ from Ref.~\cite{Zyla:2020zbs}. The coefficients $\delta_O$ of the leading chiral logarithm are known analytically \cite{Chen:2001eg,Detmold:2002nf,Green:2012ej,Wein:2014wma}
\begin{align}
 \delta_{g_A^{u-d}}  &= \delta_{\avgx{-}{\Delta}} = -\l(1+2\l(\chiral{g}_A^{u-d}\r)^2\r) \,, \\
 \delta_{g_S^{u-d}}  &= -\frac{1+6\l(\chiral{g}_A^{u-d}\r)^2}{2} \,, \\
 \delta_{g_T^{u-d}}  &= \delta_{\avgx{-}{\delta}} = -\frac{1+4\l(\chiral{g}_A^{u-d}\r)^2}{2} \,, \\
 \delta_{\avgx{-}{}} &= -\l(1+3\l(\chiral{g}_A^{u-d}\r)^2\r) \,.
\end{align}
for all six NMEs and depend only on $A_{g_A^{u-d}}=\chiral{g}_A^{u-d}$, which we treat as an additional free parameter for $O\neq g_A^{u-d}$. The $\chi$PT-part of the fit model in Eq.~(\ref{eq:NNLO_CCF_model}) has been complemented by a term $\sim D_O a^{n(O)}$ to account for the leading scaling behavior in $a$, where
\begin{equation}
 n(g_A^{u-d}) = n(g_S^{u-d}) = n(g_T^{u-d}) = 2 \,.
\end{equation}
This is in line with the implemented $O(a)$ improvement of the renormalization factors and the fact that no operator improvement is required for these NMEs at vanishing momentum transfer $Q^2=0$. On the other hand, for the NMEs associated with twist-2 operator insertions a linear behavior in $a$ is expected, i.e.
\begin{equation}
 n(\avgx{-}{}) = n(\avgx{-}{\Delta}) = n(\avgx{-}{\delta}) = 1 \,.
\end{equation}
Finally, the purpose of the last term in Eq.~(\ref{eq:NNLO_CCF_model}) is to account for finite volume effects \cite{Beane:2004rf}. \par

Besides the full NNLO model in Eq.~(\ref{eq:NNLO_CCF_model}) we consider a simpler model that is obtained by setting $C_O=\delta_O=0$, i.e. only fitting the leading light-quark mass dependence with a term $\sim M_\pi^2$
\begin{equation}
 O(M_\pi, a, L) = A_O + B_O M_\pi^2 + D_O a^{n(O)} + E_O \frac{M_\pi^2}{\sqrt{M_\pi L}} e^{-M_\pi L} \,.
 \label{eq:basic_CCF_model} 
\end{equation}
This is motivated by the generally rather flat chiral behavior of the data for
any of the NMEs. In fact, attempting to fit the full NLO expression obtained by
removing only the cubic term in $M_\pi$ leads to unacceptably large values of
$\chi^2/N_\mathrm{dof}$ in almost any case. The reason for this is that the
curvature imposed by the chiral logarithm is clearly not observed in our data.
Including the prefactor of the chiral logarithm as an independent fit parameter
in the NLO expression, one finds that the fit typically prefers the opposite
sign as predicted by $\chi$PT, which has already been observed in other studies
\cite{Ottnad:2022axz,Harris:2019bih,Capitani:2017qpc}. Fitting the full model one finds that the cubic term competes with the chiral logarithm, canceling (some of) its contribution to reproduce the rather flat behavior of the lattice data. \par

In order to further assess the stability of the CCF fits, we impose cuts to our lattice data in addition to fitting the full data sets for any of the observables. First of all, we implement a cut in the pion mass, i.e. $M_\pi<300\mev$ to test the convergence of the chiral extrapolation. Secondly, we apply a cut of $a<0.08\fm$ in the lattice spacing or a cut removing the ensembles with $M_\pi L<4$ (i.e. H105 and S201). However, either of these two cuts is only applied in combination with the cut in $M_\pi$ and not directly to the full set of data. This choice is supposed to prevent giving undue weight to the statistically more precise data at heavier pion masses in the final model averages in Sec.~\ref{sec:AIC_and_final_results}. Finally, we include an even more restrictive cut in the pion mass of $M_\pi<270\mev$ for the second CCF fit model with $C_O=\delta_O=0$ to further scrutinize the chiral extrapolation. However, this cut cannot be applied in case of the full NNLO model as it leaves only six data points, which is not enough to constrain the six (five in case of $g_A^{u-d}$) fit parameters of this model. \par

\subsection{Nucleon charges} \label{subsec:CCF_charges}

\begin{figure}[!htb]
 \centering
 \includegraphics[totalheight=0.226\textheight]{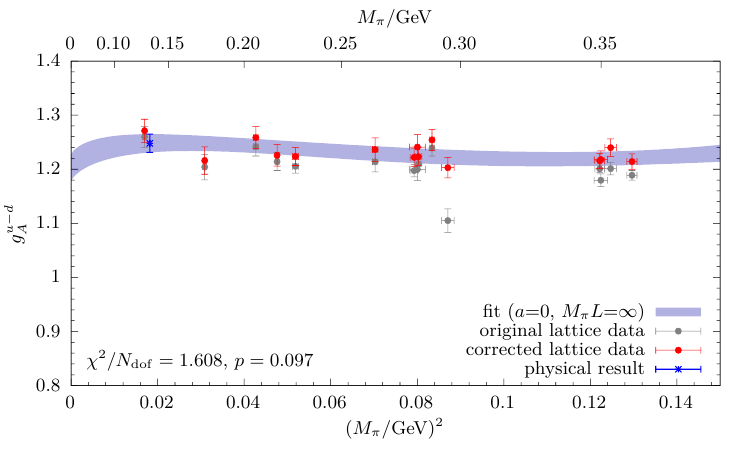}
 \includegraphics[totalheight=0.226\textheight]{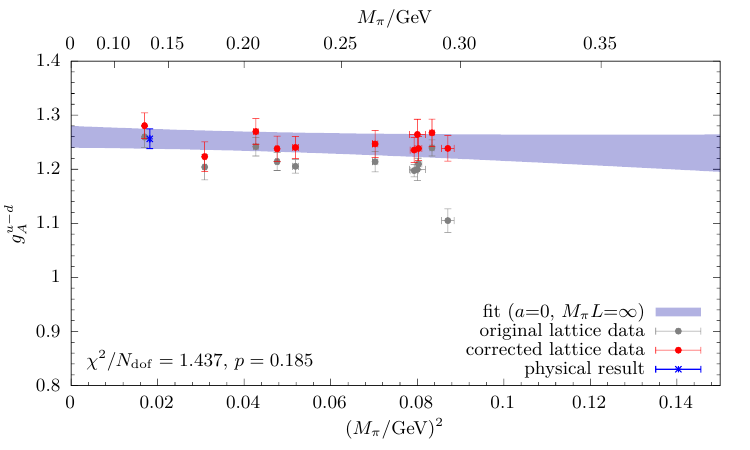} \\
 \includegraphics[totalheight=0.226\textheight]{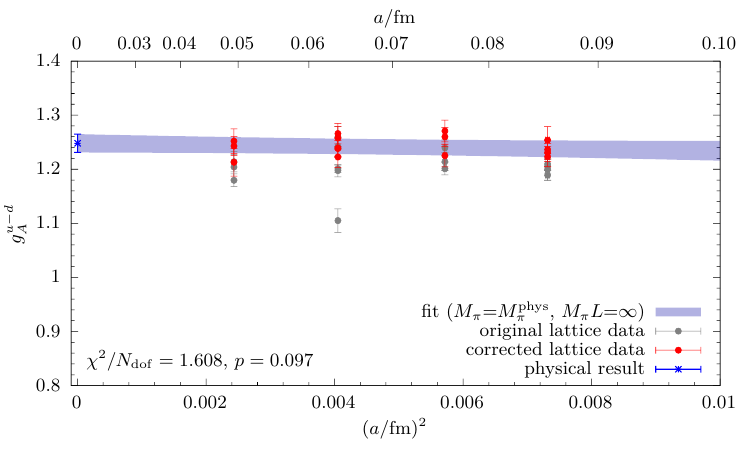}
 \raisebox{-0.58em}{\includegraphics[totalheight=0.226\textheight]{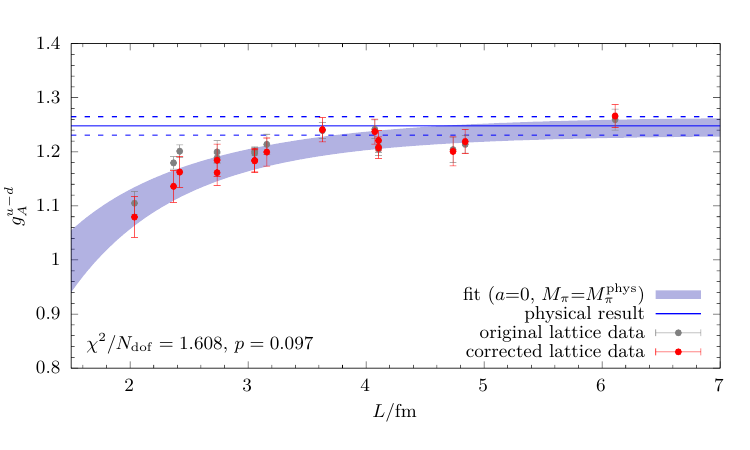}}
 \caption{Examples for the physical extrapolation of $g_A^{u-d}$ using lattice data from the two-state summation fit model ansatz in Eq.~(\ref{eq:summation_two_state}) with $\tsepmin\approx0.3\fm$. Upper row: chiral extrapolation as a function of $M_\pi^2$ fitting the full set of data to the NNLO model in Eq.~(\ref{eq:NNLO_CCF_model}) (left panel) and with a cut in $M_\pi<300\mev$ using the simplified model in Eq.~(\ref{eq:basic_CCF_model}) (right panel). Lower row: continuum (left panel) and infinite volume (right panel) extrapolations fitting the full set of data to the NNLO fit model in Eq.~(\ref{eq:NNLO_CCF_model}). The red data points are obtained by correcting the original lattice data for the extrapolations in all variables but the one on the x-axis, using the parameters from the fit. Therefore, the resulting point errors are highly correlated. Errors are statistical only.}
 \label{fig:CCF_g_A}
\end{figure}

\begin{figure}[!htb]
 \centering
 \includegraphics[totalheight=0.226\textheight]{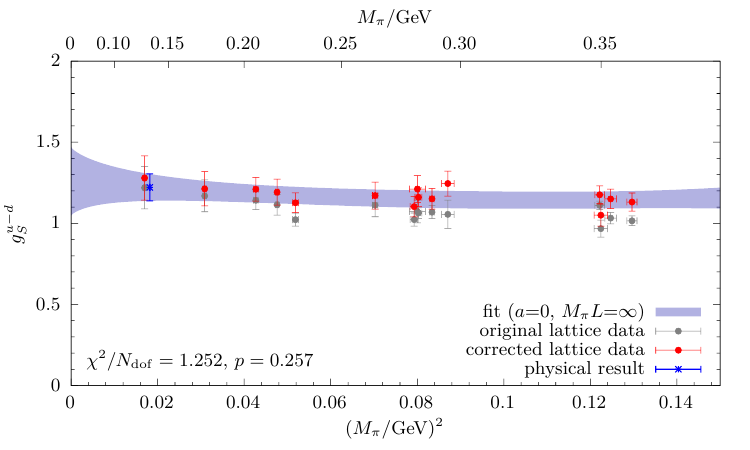}
 \includegraphics[totalheight=0.226\textheight]{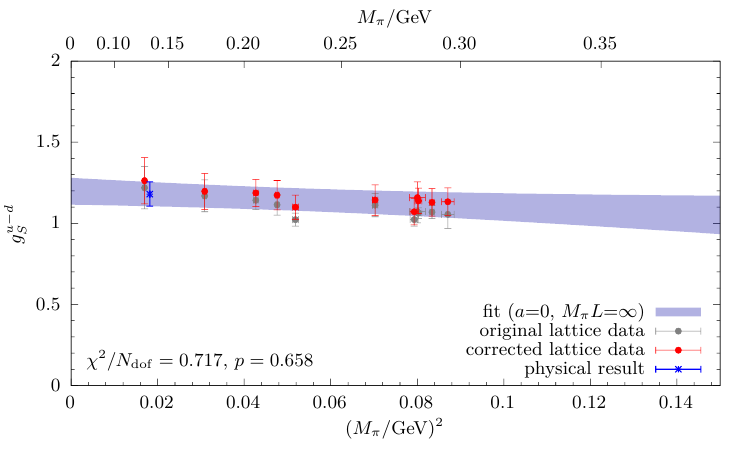}
 \includegraphics[totalheight=0.226\textheight]{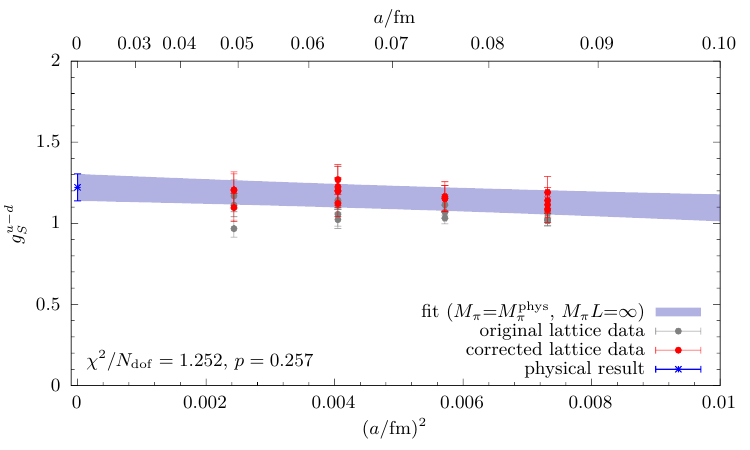}
 \raisebox{-0.58em}{\includegraphics[totalheight=0.226\textheight]{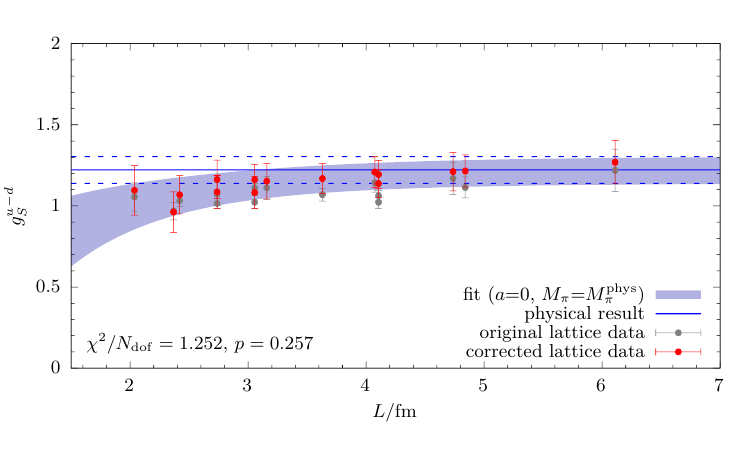}}
 \vspace*{-1.0em}
 \caption{Same as Fig.~\ref{fig:CCF_g_A} but for $g_S^{u-d}$.}
 \label{fig:CCF_g_S}
\end{figure}

\begin{figure}[!htb]
 \centering
 \includegraphics[totalheight=0.226\textheight]{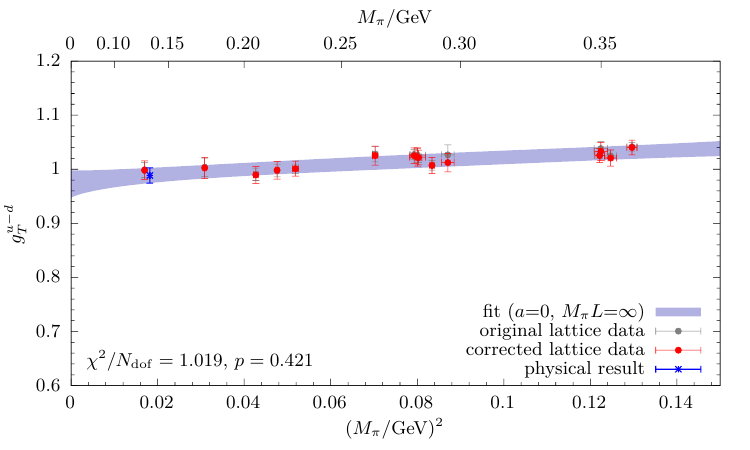}
 \includegraphics[totalheight=0.226\textheight]{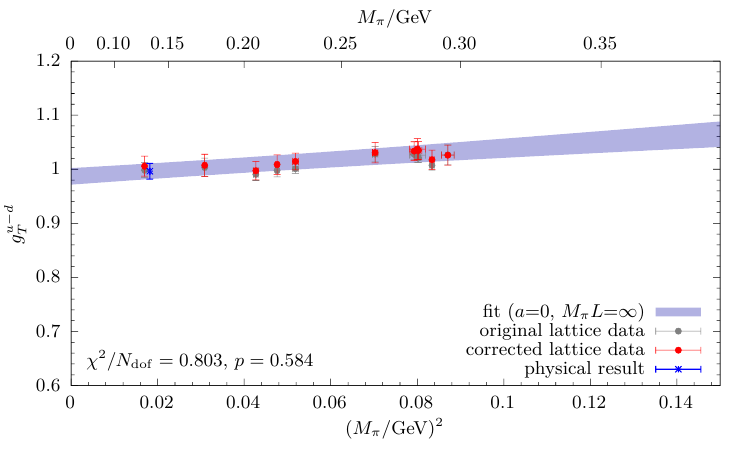}
 \includegraphics[totalheight=0.226\textheight]{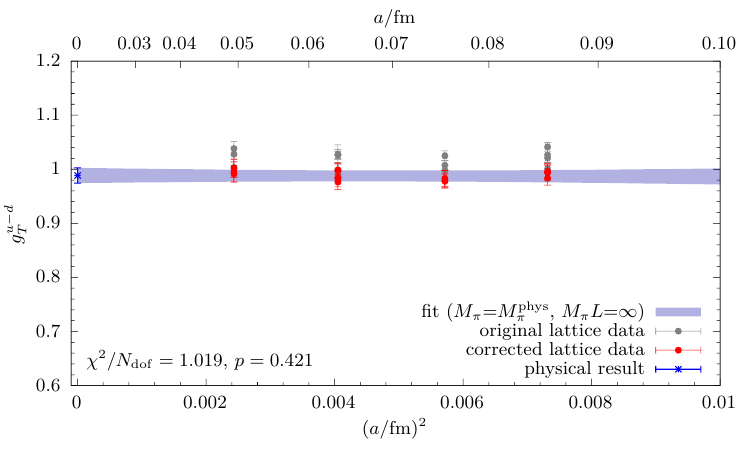}
 \raisebox{-0.58em}{\includegraphics[totalheight=0.226\textheight]{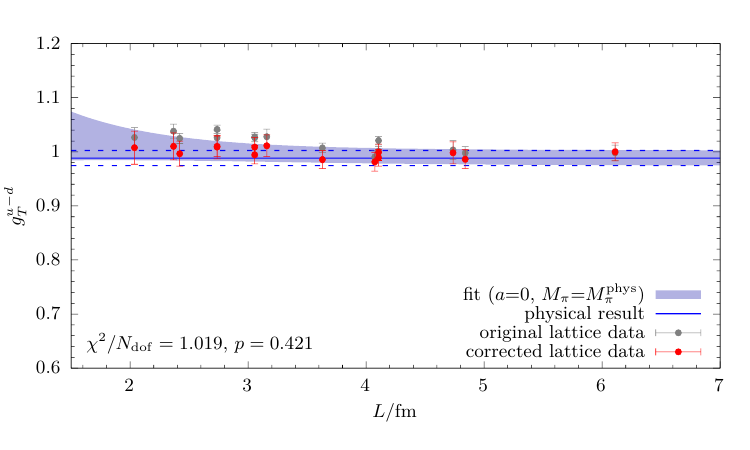}}
 \vspace*{-1.0em}
 \caption{Same as Fig.~\ref{fig:CCF_g_A} but for $g_T^{u-d}$.}
 \label{fig:CCF_g_T}
\end{figure}

In Figs.~\ref{fig:CCF_g_A},~\ref{fig:CCF_g_S}~and~\ref{fig:CCF_g_T} results are shown for the physical extrapolation of $g_A^{u-d}$, $g_S^{u-d}$ and $g_T^{u-d}$, respectively. The chiral extrapolation is found to be rather mild: for $g_A^{u-d}$ and $g_S^{u-d}$ it is almost flat, while for $g_T^{u-d}$ the fit prefers a positive slope in $M_\pi^2$. The latter yields a correction of at most $\lesssim 5\%$ for ensembles at around $M_\pi \approx 350\mev$ towards the physical result for $g_T^{u-d}$. The full set of data is generally described well by the NNLO model in Eq.~(\ref{eq:NNLO_CCF_model}). In particular for the statistically most precise data for $g_T^{u-d}$ the resulting values of $\chi^2/N_\mathrm{dof}$ indicate an excellent description of the data by the fit model(s). The same holds true for fits of the more simplistic fit model in Eq.~(\ref{eq:basic_CCF_model}). A data cut of $M_\pi<300\mev$ as shown in the upper right panels of Figs.~\ref{fig:CCF_g_A}--\ref{fig:CCF_g_T} may further reduce $\chi^2/N_\mathrm{dof}$ for fits to Eq.~(\ref{eq:basic_CCF_model}) compared to fitting the full dataset to the same model. The fact that the more sophisticated model in Eq.~(\ref{eq:NNLO_CCF_model}) does not necessarily yield a better description of the data, as observed for e.g. $g_A^{u-d}$, can be attributed to the aforementioned issue concerning the sign of the chiral logarithm. For example, fitting Eq.~(\ref{eq:NNLO_CCF_model}) without the cubic term (i.e. setting $C_{g_A^{u-d}}=0$) to either the full dataset or the one with $M_\pi<300\mev$ that have been used for the upper two panels of Fig.~\ref{fig:CCF_g_A} results in $\chi^2/N_\mathrm{dof}=5.072$ and $\chi^2/N_\mathrm{dof}=3.299$, respectively. Since such fit models based on NLO $\chi$PT clearly fail to describe the data, we do not further consider them in our final analysis. Besides, they would carry essentially zero weight in a model average. \par

Examples for the continuum extrapolation for the $\mathcal{O}(a)$-improved local isovector nucleon charges are displayed in the lower left panels of Figs.~\ref{fig:CCF_g_A}--\ref{fig:CCF_g_T}. It is basically found to be compatible with a constant for all three observables. While fit results for $D_O$ may show some variation depending on the choice of the fit model and data cuts, the correction due to the continuum limit falls typically within the statistical errors of the physical result, indicating that systematic effects due to the continuum extrapolation are indeed well under control. \par

The observed model independence of the chiral extrapolation and the flat continuum extrapolation, taken together with the fact that the chirally extrapolated values are in very good statistical agreement with the results obtained on our (slightly lighter than) physical pion-mass ensemble E250, indicate that our physical results are not strongly reliant on the validity of chiral perturbation theory or the Symanzik effective theory.

Finally, the extrapolations to infinite volume that are shown in the lower right panels of Figs.~\ref{fig:CCF_g_A}--\ref{fig:CCF_g_T} exhibit a rather peculiar pattern. On the one hand, the extrapolation in $L$ is found to be entirely flat for the very precise data for $g_T^{u-d}$ and very stable under any cuts that are applied to the data. On the other hand, one finds large corrections for $g_A^{u-d}$, an observation that was already made in Ref.~\cite{Harris:2019bih}. On the ensemble S201 with the smallest volume ($L\approx 2\fm$), the correction exceeds $10\%$ and at $L=3\fm$ it is still well around the $5\%$-level. It is only for ensembles with $L>4\fm$ that the finite volume correction for $g_A^{u-d}$ starts falling below the statistical error of the final results of the fit. This feature remains qualitatively the same even when applying a cut of $M_\pi L\gtrsim4$ to the lattice data entering the CCF fit, i.e. the fit still resolves the curvature albeit with larger errors.  At any rate, it is reassuring to observe that the most chiral ensemble E250, which also exhibits the largest physical volume corresponding to $L\approx 6.1\fm$, clearly confirms the result of the infinite-volume extrapolation within the statistical accuracy. \par


\subsection{Twist-2 matrix elements} \label{subsec:CCF_twist2_NMEs}
\begin{figure}[!htb]
 \centering
 \includegraphics[totalheight=0.226\textheight]{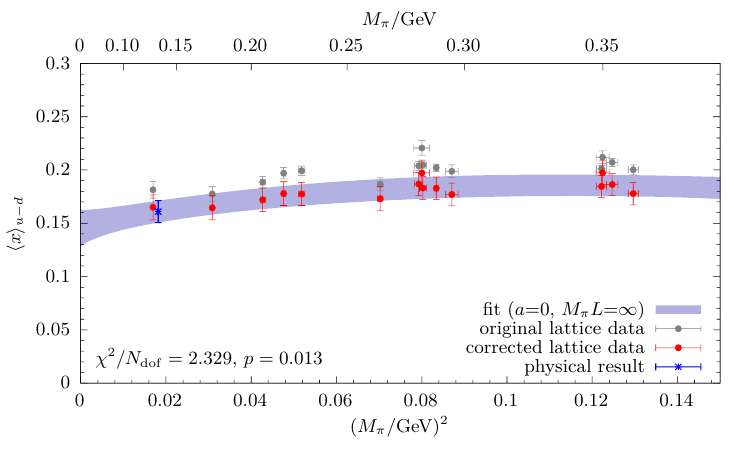}
 \includegraphics[totalheight=0.226\textheight]{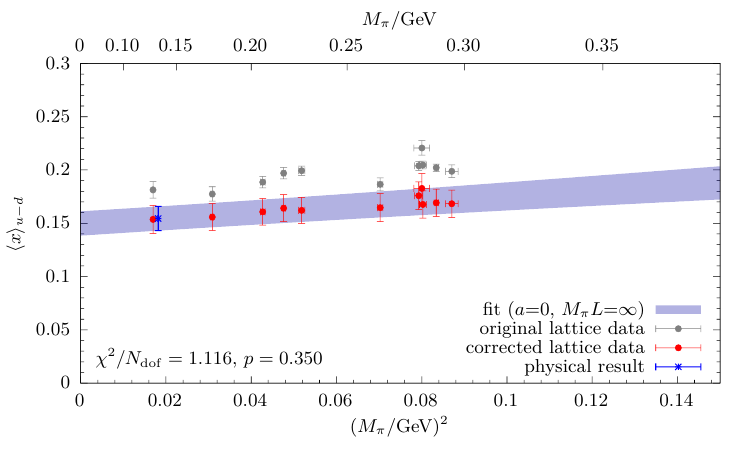} \\
 \includegraphics[totalheight=0.226\textheight]{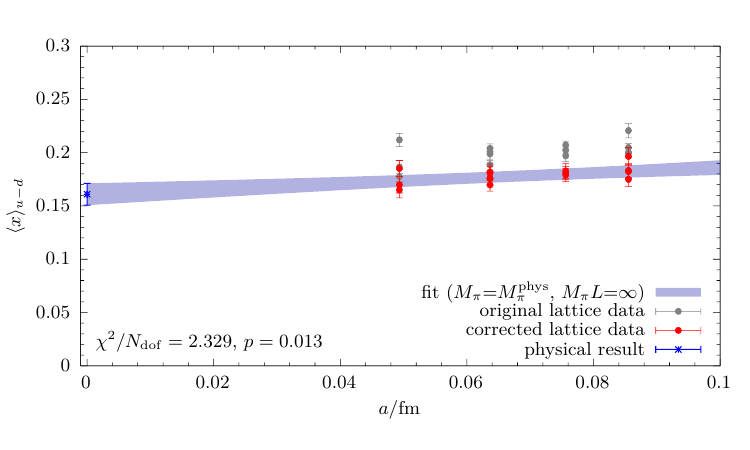}
 \includegraphics[totalheight=0.226\textheight]{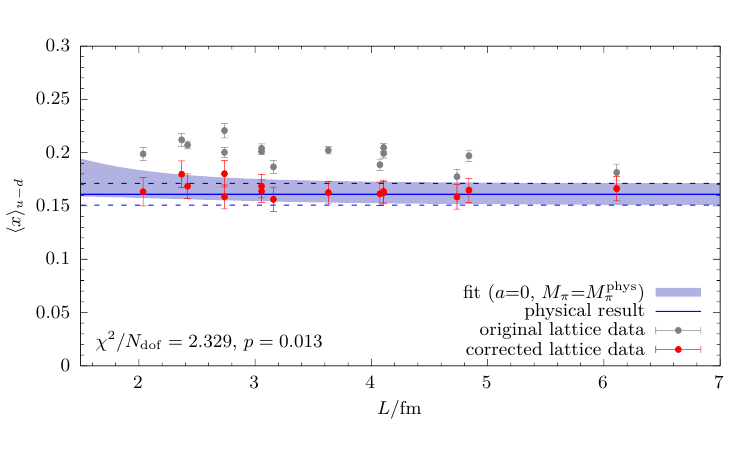}
 \caption{Same as Fig.~\ref{fig:CCF_g_A} but for $\avgx{-}{}$.}
 \label{fig:CCF_avgx}
\end{figure}

\begin{figure}[!htb]
 \centering
 \includegraphics[totalheight=0.226\textheight]{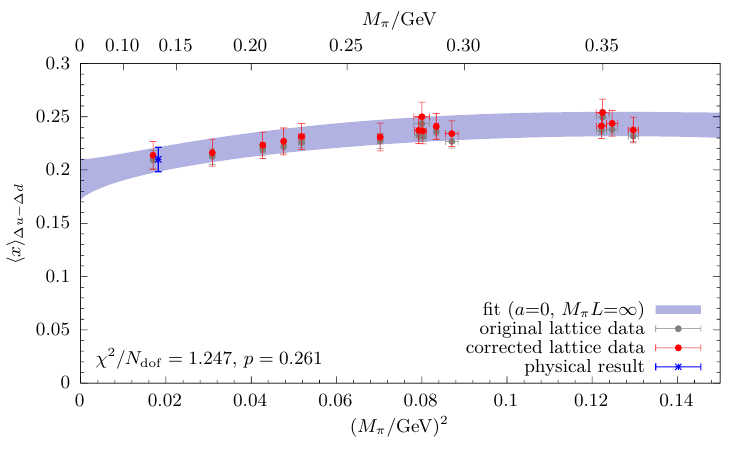}
 \includegraphics[totalheight=0.226\textheight]{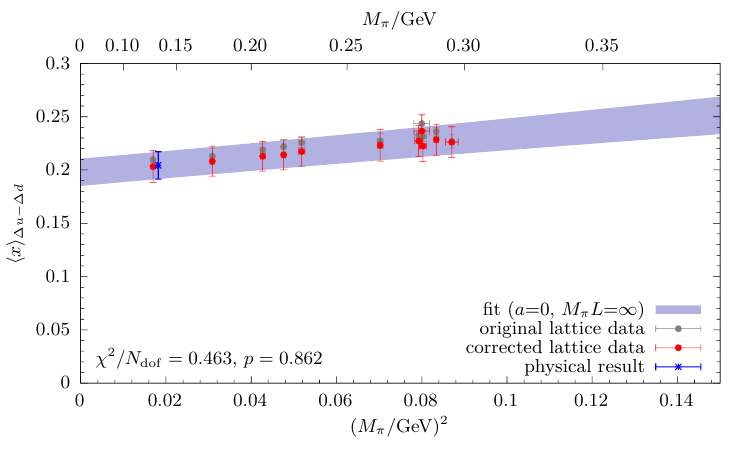} \\
 \includegraphics[totalheight=0.226\textheight]{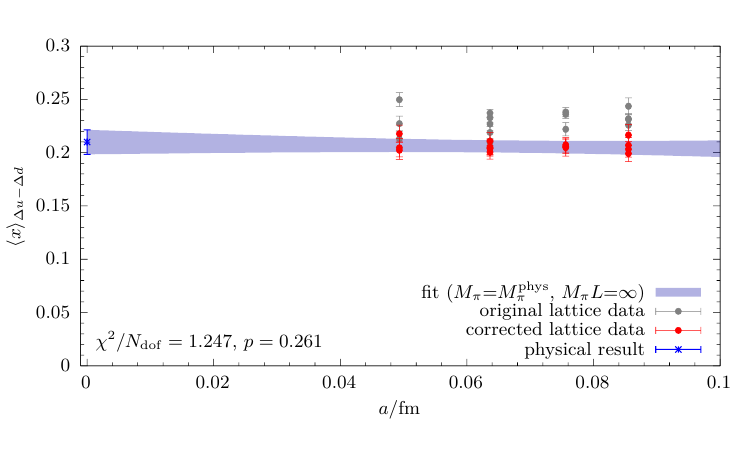}
 \includegraphics[totalheight=0.226\textheight]{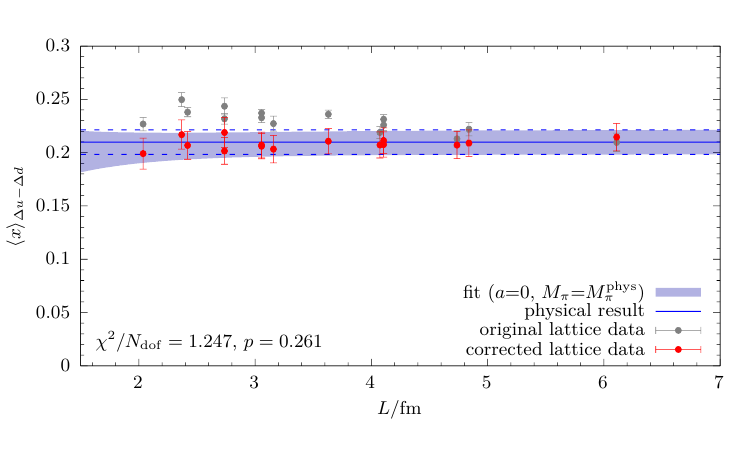}
 \caption{Same as Fig.~\ref{fig:CCF_g_A} but for $\avgx{-}{\Delta}$.}
 \label{fig:CCF_helicity}
\end{figure}

\begin{figure}[!htb]
 \centering
 \includegraphics[totalheight=0.226\textheight]{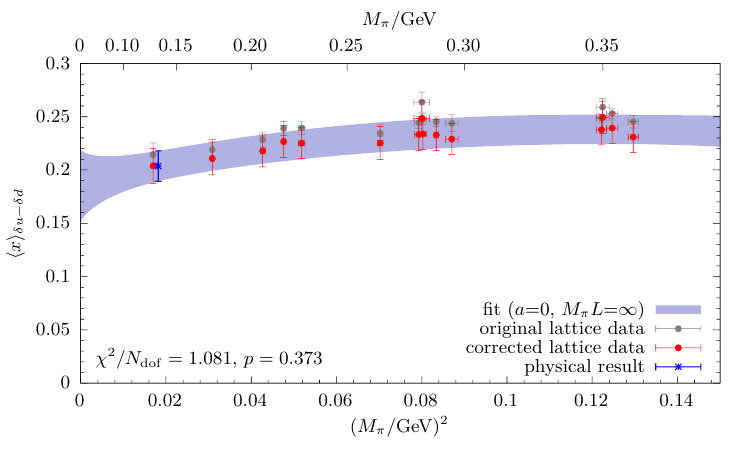}
 \includegraphics[totalheight=0.226\textheight]{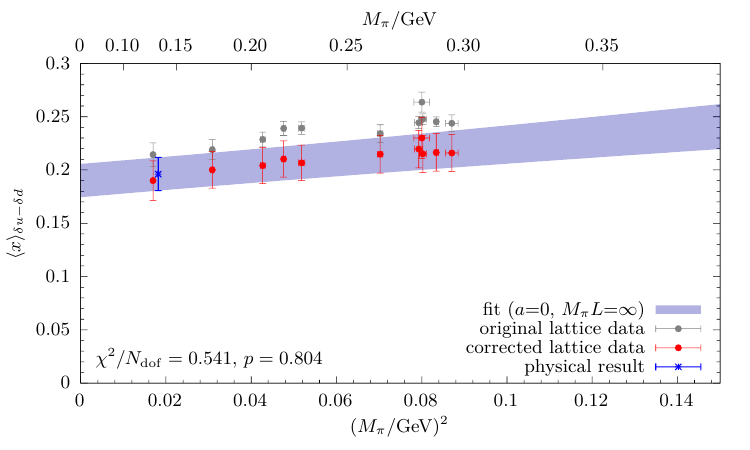} \\
 \includegraphics[totalheight=0.226\textheight]{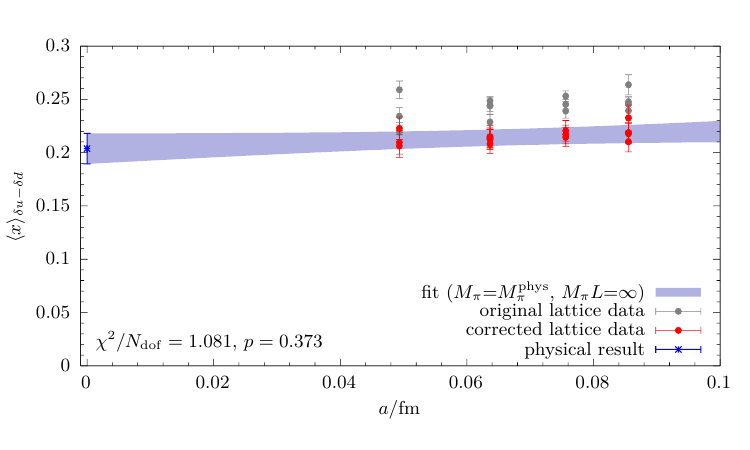}
 \includegraphics[totalheight=0.226\textheight]{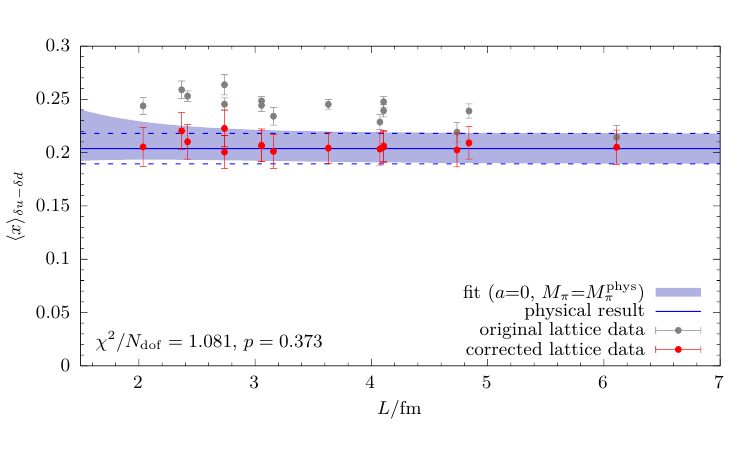}
 \caption{Same as Fig.~\ref{fig:CCF_g_A} but for $\avgx{-}{\delta}$.}
 \label{fig:CCF_transversity}
\end{figure}

Results for the physical extrapolations of the twist-2 isovector NMEs are displayed in Figs.~\ref{fig:CCF_avgx},~\ref{fig:CCF_helicity}~and~\ref{fig:CCF_transversity}. The most striking feature of these extrapolations is the observed similarity between the three different operators insertions, i.e. their chiral behavior is always characterized by a positive slope and a very similar curvature. Corrections towards the physical point limit due to the chiral extrapolation roughly reach the $\sim 10\%$ level for ensembles at the light quark masses corresponding to $M_\pi\approx 250\mev$. Generally, a fit of the NNLO model in Eq.~(\ref{eq:NNLO_CCF_model}) to the full set of data for the twist-2 NMEs results in a more pronounced nonlinear curvature as a function of $M_\pi^2$ than for the local charges. However, the chiral behavior of the twist-2 NMEs is also not incompatible with a linear extrapolation in $M_\pi^2$. This is particularly true when applying a cut in $M_\pi$ as shown in the upper right panels of Figs.~\ref{fig:CCF_avgx}--\ref{fig:CCF_transversity}. Still, for the full set of data including ensembles with $M_\pi>300\mev$ the simplified model in Eq.~(\ref{eq:basic_CCF_model}) leads to worse $p$-values as compared to fitting the NNLO model in Eq.~(\ref{eq:NNLO_CCF_model}), i.e. $p=0.005$, $p=0.106$ and $p=0.207$ for $\avgx{-}{}$, $\avgx{-}{\Delta}$ and $\avgx{-}{\delta}$, respectively. Again, the physical results for the twist-2 NMEs are found to be in good agreement with the results of the old analysis with significantly reduced statistical errors. We note that for all three twist-2 NMEs there is also broad agreement for the ground-state NMEs data on the individual ensembles that enter the CCF fits between the old and the new analysis for the common subset of ensembles. The statistical error for the NMEs on the individual ensembles is typically reduced by a factor $\sim2$ to $\sim10$ in the present study. \par

The continuum extrapolation is reasonably flat for all three NMEs. Variations due to the choice of the fit ansatz and data cuts are fairly mild and the results for the fit parameter $D_O$ are typically compatible with zero within $\lesssim 2\sigma$. We observe a preference for a positive slope as a function of $a$ for $\avgx{-}{}$ and $\avgx{-}{\delta}$ across the various fits, whereas for $\avgx{-}{\Delta}$ there is no significant trend at all. The corrections due to the continuum limit are typically $\lesssim 10\%$ of the physical result and thus within $1$--$2\sigma$ of its central value. This can be seen as a rather strong indication that also for the twist-2 NMEs the continuum limit is not a major source of systematic error, despite the lack of $\mathcal{O}(a)$ improvement. \par

Finally, the infinite-volume extrapolation as a function of $L$ in the lower right panels of Figs.~\ref{fig:CCF_avgx}--\ref{fig:CCF_transversity} is almost entirely flat, and there is no clear preference for the sign of the fit coefficient $E_O$ across different fit models and data cuts. In fact, the fit coefficient $E_O$ is also found to be well compatible with zero for the majority of fits. Finally, we note that the individual, continuum corrected result for the twist-2 NMEs on the E250 ensemble with the largest physical volume are generally in very good agreement with the trend of the infinite volume extrapolations (as well as with the resulting physical results themselves). \par

As for the nucleon charges, we observe no significant dependence of the twist-2 matrix elements on the form of the chiral extrapolation, and the result at the physical point agrees with the result on E250 within statistical errors.

\section{Model averages and final results} \label{sec:AIC_and_final_results}
The results from individual CCF fits for any of the isovector NMEs and choice of $\tsepmin\in{0.2\fm,0.3\fm,0.4\fm}$ are combined in a model average based on a variation of the Akaike information criterion (AIC) \cite{1100705,Akaike1998}. To this end, we assign weights \cite{doi:10.1177/0049124104268644,BMW:2014pzb,Neil:2022joj}
\begin{equation}
  w_{n,b} = \frac{e^{-B_{n,b}/2}}{\sum_{k=1}^{N_M} e^{-B_{k,b} / 2}} \,,
  \label{}
\end{equation}
to each model with index $n\in{1,...,N_M}$ on every bootstrap sample $b\in{1,...,N_B}$, where
\begin{equation}
  B_{n,b} = \chi^2_{n,b} + 2 N_{\mathrm{par},n} + 2 N_{\mathrm{cut},n} \,,
  \label{eq:BAIC}
\end{equation}
is the Bayesian AIC introduced in Ref.~\cite{Neil:2022joj}. In this expression $\chi^2_{n,b}$ refers to the minimized, correlated $\chi^2$ from the $n$-th fit model on the $b$-th bootstrap sample, and $N_{\mathrm{par},n}$, $N_{\mathrm{cut},n}$ denote the numbers of fit parameters and cut data points in the corresponding model, respectively. Note that there are no priors used in any of the CCF models, hence there is no need to account for them in the computation of $B_n$. In order to disentangle the statistical and systematic contribution to the final errors, we employ a procedure similar to the one introduced in Ref.~\cite{Borsanyi:2020mff}. However, in the definition of the cumulative distribution function (CDF) for the model-averaged observables 
\begin{equation}
  CDF(y)= \frac{1}{N_B} \sum_{n=1}^{N_M} \sum_{b=1}^{N_B} w_{n,b} \Theta(y-O_{n,b}) \,,
  \label{eq:CDF}
\end{equation}
we make direct use of the actual bootstrap distributions rather than assuming (weighted) Gaussian CDFs for the individual models that are constructed from central values and errors as in Ref.~\cite{Borsanyi:2020mff}. In this expression, $\Theta$ is the Heaviside step function, and the outer sum runs over $N_M=27$ different models, whereas the inner sum encompasses bootstrap results for every observable $O_{n,b}$ for any given model $n$. The central value and total error of our final, physical results are given by the median and the quantiles corresponding to $1\sigma$ errors for a Gaussian distribution, respectively. Statistical and systematic contributions to the total error are then determined in a similar way as in Ref.~\cite{Borsanyi:2020mff}, i.e. by rescaling the statistical error, which in our case corresponds to rescaling the width of the individual bootstrap distributions. For this purpose we employ the same choice of $\lambda=2$ for the rescaling factor as in Ref.~\cite{Borsanyi:2020mff}. We remark that results for the errors are virtually independent of the choice of $\lambda$ when $\lambda \geq 2$. \par

\begin{figure}[t]
 \centering
  \includegraphics[totalheight=0.226\textheight]{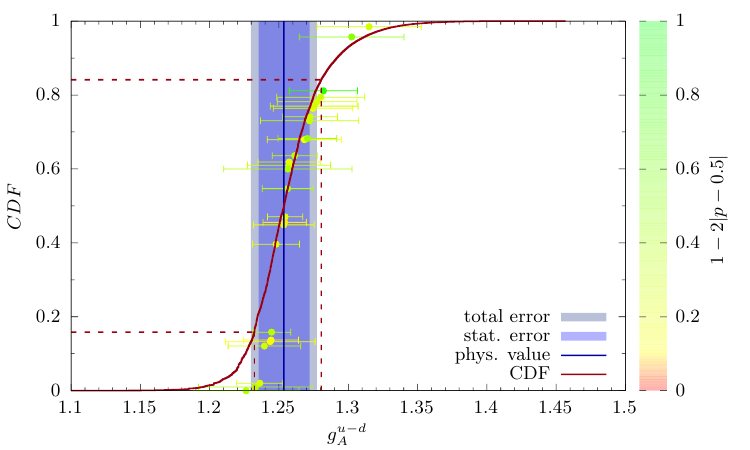}
  \includegraphics[totalheight=0.226\textheight]{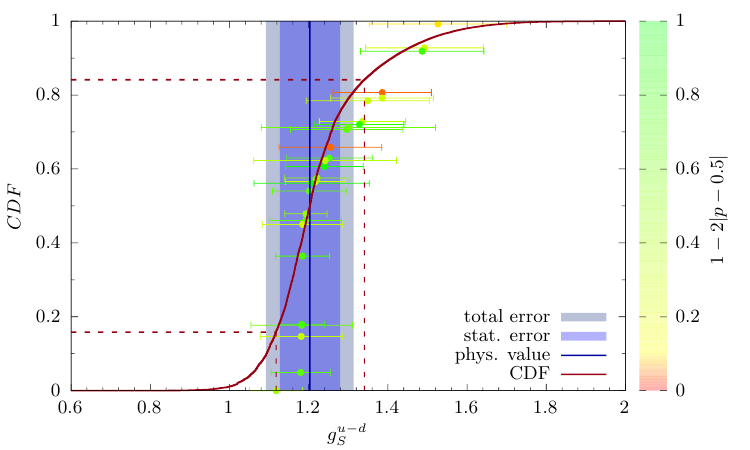} \\
  \includegraphics[totalheight=0.226\textheight]{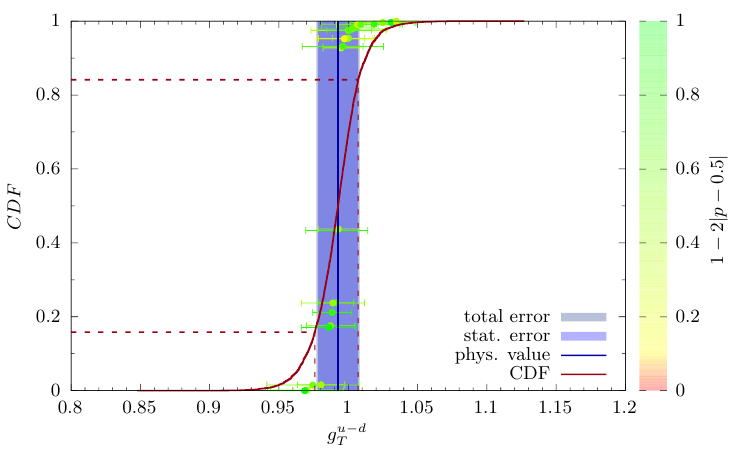}
  \includegraphics[totalheight=0.226\textheight]{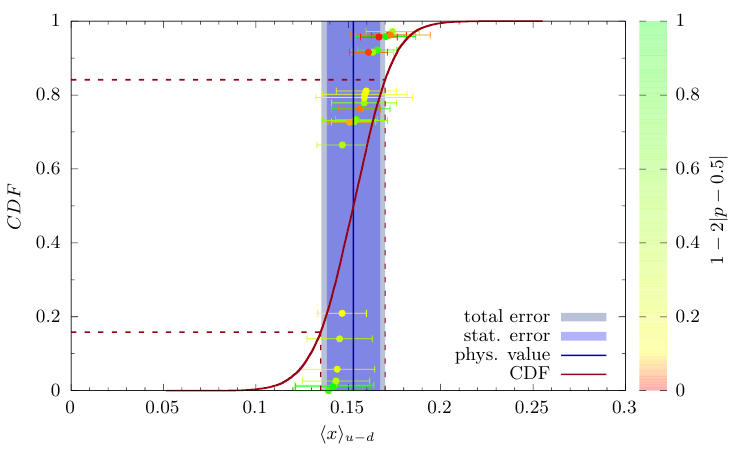} \\
  \includegraphics[totalheight=0.226\textheight]{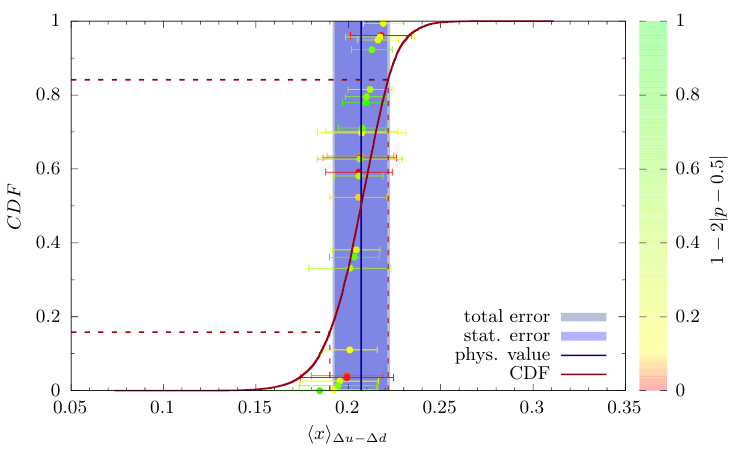}
  \includegraphics[totalheight=0.226\textheight]{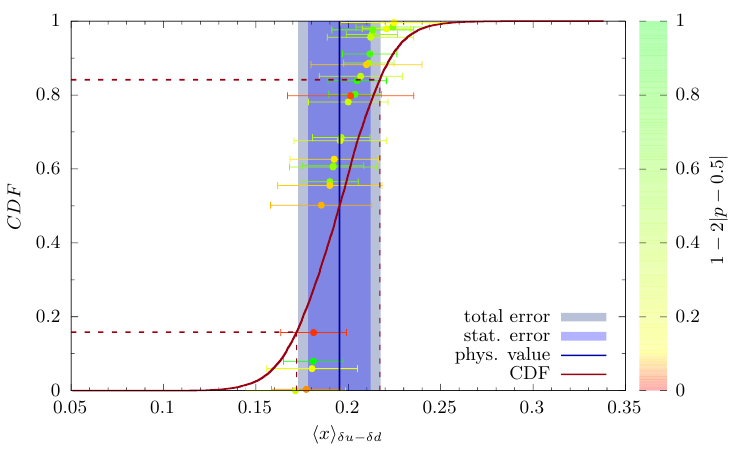}
  \caption{Cumulative distribution functions (CDFs) of the fit models for all six isovector NMEs. Results of the fit models with statistical errors are represented by the individual data points. The color of these data points is referring to the p-value weight which is different from the Akaike weight used in the actual CDF to allow for a visual assessment of the quality of the individual fits. The final result from the model average is given by the solid blue line together with its symmetrized statistical and full error bands as indicated in the plots. The dashed lines represent the (generally non-symmetric) $1\sigma$-quantiles of the CDF.}
  \label{fig:AIC_final_results}
\end{figure}

Results from this procedure are shown in Fig.~\ref{fig:AIC_final_results} and the final, physical results with statistical and systematic errors from the model averaging read 
\begin{align}                                                          
 g_A^{u-d} &= 1.254\stat{19}\sys{15}\total{24},  \label{eq:gA_phys} \\ 
 g_S^{u-d} &= 1.203\stat{77}\sys{81}\total{112}, \label{eq:gS_phys} \\ 
 g_T^{u-d} &= 0.993\stat{15}\sys{05}\total{16},  \label{eq:gT_phys}    
\end{align}
for the isovector NMEs of local operator insertions, and
\begin{align}                                                                  
 \avgx{-}{}       &= 0.153\stat{15}\sys{10}\total{17}, \label{eq:avgx_phys} \\ 
 \avgx{-}{\Delta} &= 0.207\stat{15}\sys{06}\total{16}, \label{eq:helx_phys} \\ 
 \avgx{-}{\delta} &= 0.195\stat{17}\sys{15}\total{23}, \label{eq:trvx_phys}    
\end{align}
for the isovector moments of twist-2 operator insertions, respectively. The systematic error reflects the combined uncertainties associated with the chiral, continuum and infinite volume limits in the physical extrapolation, as well as the uncertainty due to the choice of $\tsepmin$ in the determination of the ground-state NME values that enter these final fits. Overall, our results show a good balance between statistical and systematic errors.\par

\section{Comparison and Outlook}
\label{sec:summary}

The chiral extrapolations and physical results for the three local charges are in broad agreement with the results of our earlier analysis on a subset of the ensembles with $M_\pi\gtrsim 200\mev$ in Ref.~\cite{Harris:2019bih}. However, results for the ground-state NMEs on the individual ensembles that enter the CCF fits do not always agree with the corresponding results of the older analysis on the common subset of ensembles in Table~VI of Ref.~\cite{Harris:2019bih}. In particular, for $g_A^{u-d}$ there is a trend towards larger values in the present study with differences of up to a few percent on some of the ensembles, that are not consistently covered by the larger statistical errors of the old study. This can be attributed to residual excited state contamination, which generally lead to smaller values for $g_A^{u-d}$, and gives an indication that the fit ansatz based on the NLO summation method indeed yields superior suppression of excited states at least for $g_A^{u-d}$ than the two-state fits to the ratio data as defined in Eq.~(\ref{eq:ratio_fit_two_state}) that were used in Ref.~\cite{Harris:2019bih} (where the summation method only served as a crosscheck within its significantly larger errors at that time). For the even more precise data for $g_T^{u-d}$ we do not observe such a systematic trend, and similarly for $g_S^{u-d}$ no clear trend is seen within the much larger statistical uncertainties of the old analysis. \par

The RQCD collaboration has obtained results for the local charges of the octet
baryons on a partially overlapping set of gauge ensembles, quoting
$g_A^{u-d}=1.284\genfrac{}{}{0pt}{1}{+28}{-27}$, $g_S^{u-d}=1.11\genfrac{}{}{0pt}{1}{+14}{16}$, and
$g_T^{u-d}=0.984\genfrac{}{}{0pt}{1}{+19}{-29}$ for the isovector charges of the nucleon \cite{Bali:2023sdi}.
Overall statistics of our present study are higher, as is reflected in our
smaller overall errors, while otherwise there is good agreement to within the
quoted uncertainties.

\begin{figure}[t]
 \centering
  \includegraphics[width=0.49\textwidth,keepaspectratio=]{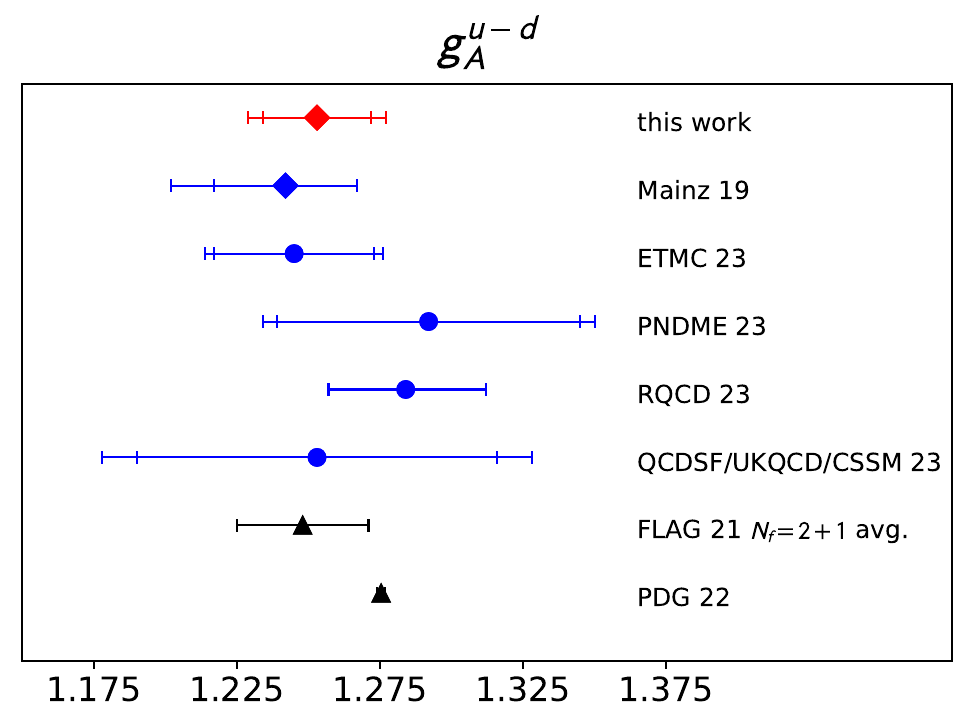}
  \includegraphics[width=0.49\textwidth,keepaspectratio=]{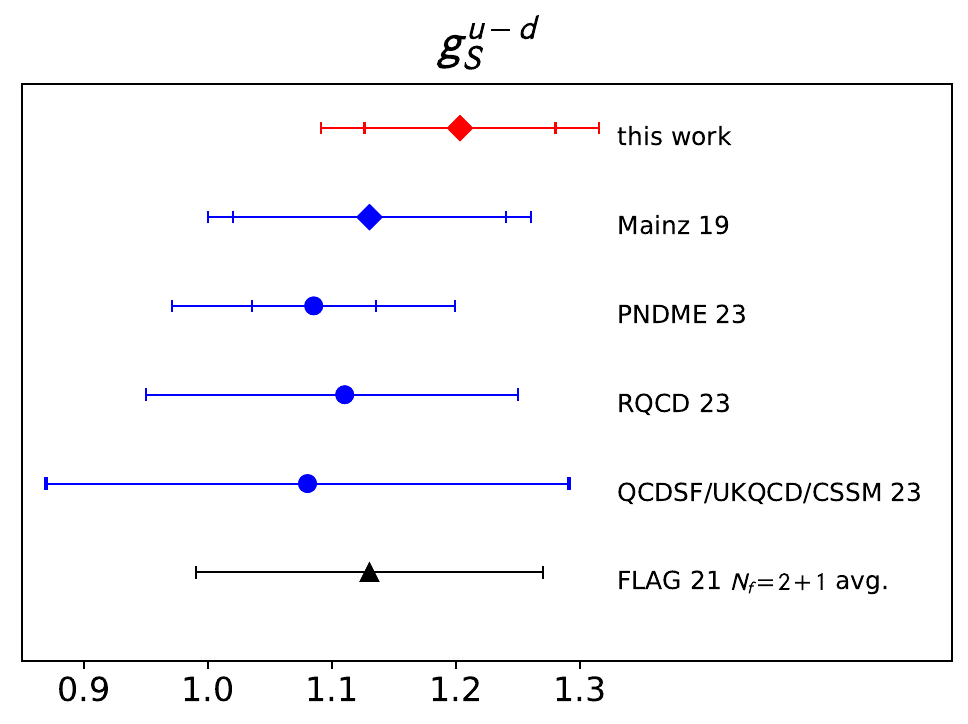} \\
  \includegraphics[width=0.49\textwidth,keepaspectratio=]{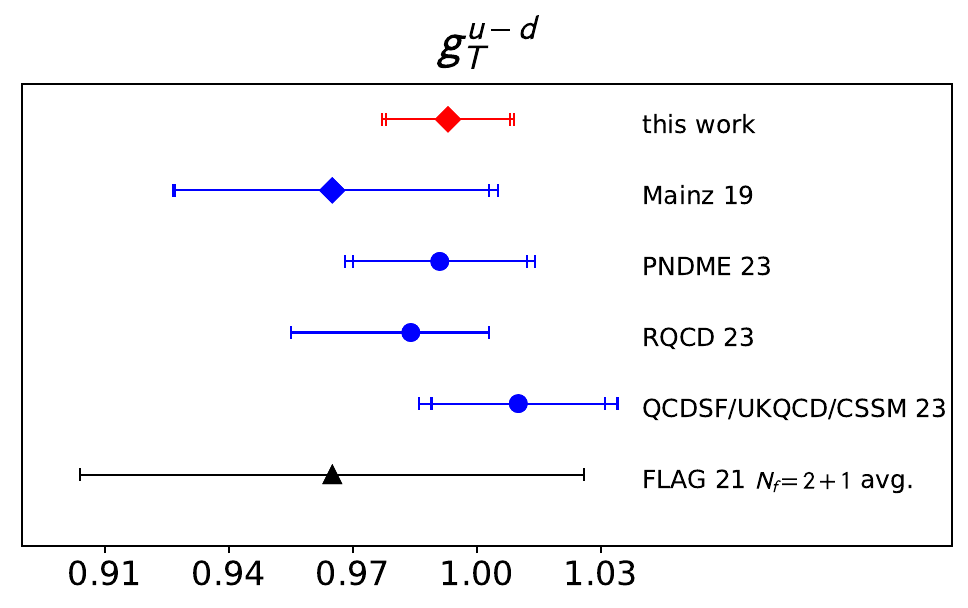}
  \includegraphics[width=0.49\textwidth,keepaspectratio=]{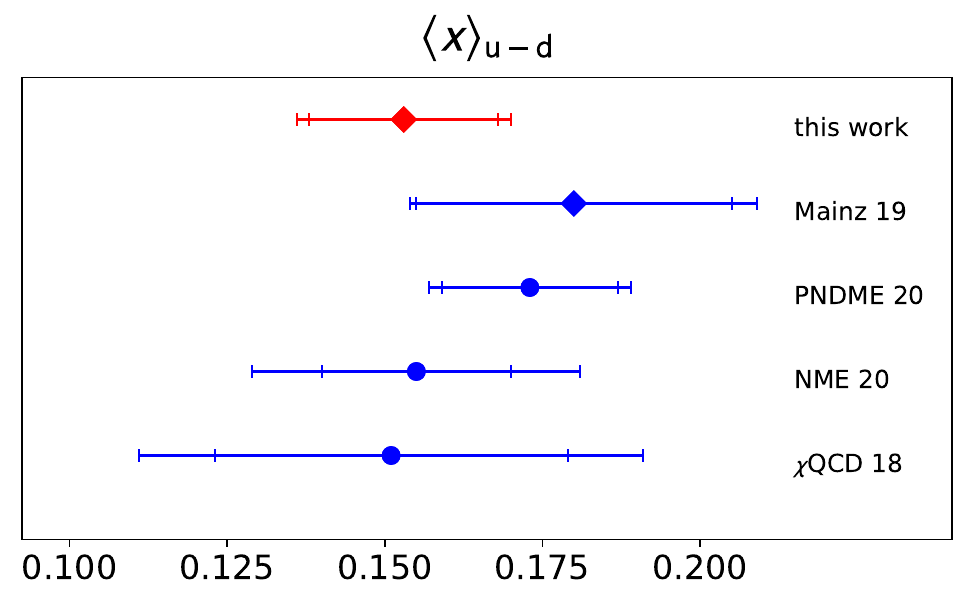} \\
  \includegraphics[width=0.49\textwidth,keepaspectratio=]{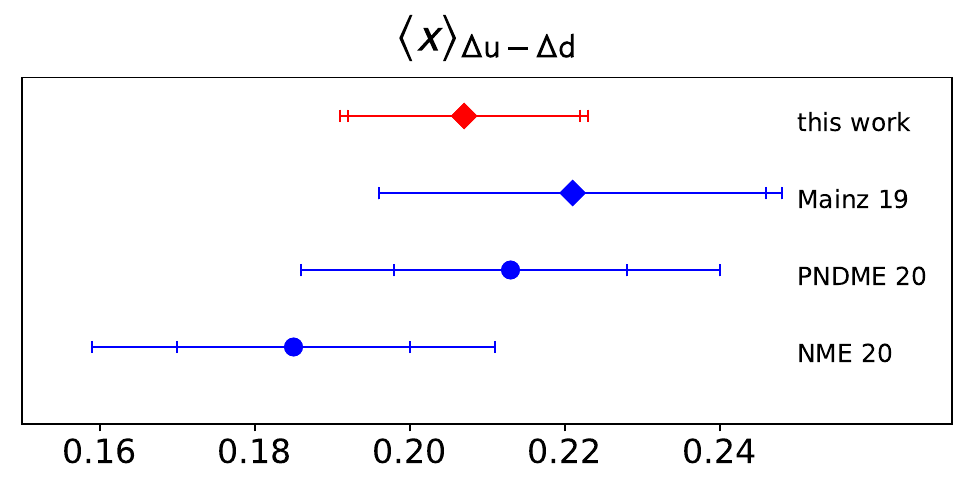}
  \includegraphics[width=0.49\textwidth,keepaspectratio=]{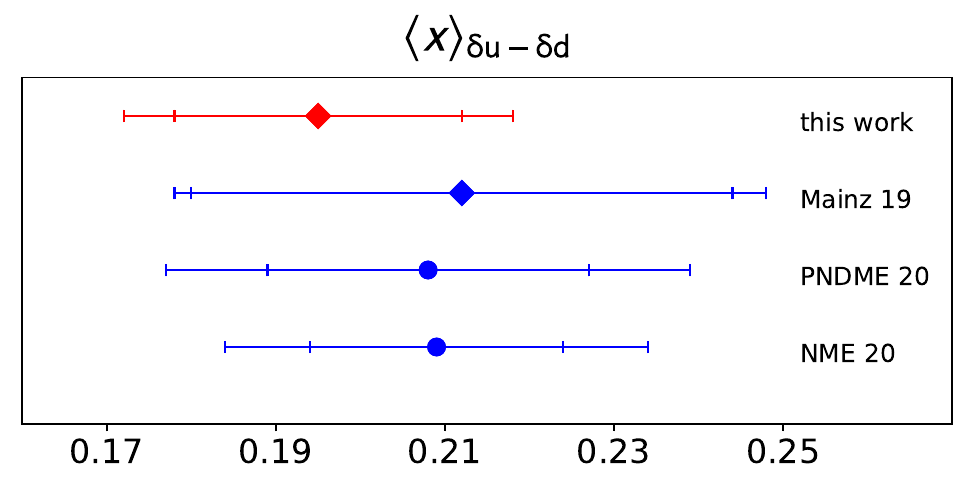}
  \caption{Comparison of our results (red diamonds) to our 2019 paper (Mainz 19
\cite{Harris:2019bih}), to recent other studies
(blue circles: QCDSF/UKQCD/CSSM 23 \cite{QCDSFUKQCDCSSM:2023qlx}, RQCD 23 \cite{Bali:2023sdi},
PNDME 23 \cite{Jang:2023zts}, ETMC 23 \cite{Alexandrou:2023qbg},
PNDME 20 \cite{Mondal:2020cmt}, NME 20 \cite{Mondal:2020ela},
$\chi$QCD 18 \cite{Yang:2018nqn})
and to the FLAG\,2021 \cite{FlavourLatticeAveragingGroupFLAG:2021npn} and 
PDG \cite{ParticleDataGroup:2022pth} averages
(black triangles, where available) for all six isovector NMEs.
Studies that entered the FLAG average are not shown separately.
Inner error bars are statistical errors only,
outer error bars include systematic errors added in quadrature. }
  \label{fig:comparisons}
\end{figure}

In Fig.~\ref{fig:comparisons}, we compare our results to other recent
determinations
\cite{QCDSFUKQCDCSSM:2023qlx,Bali:2023sdi,Jang:2023zts,Alexandrou:2023qbg,Mondal:2020cmt,Mondal:2020ela,Yang:2018nqn} 
of the isovector nucleon matrix elements, as well as
to the FLAG\,2021 averages \cite{FlavourLatticeAveragingGroupFLAG:2021npn}
in the case of the local charges. For
$g_A^{u-d}$, we also show the PDG value \cite{ParticleDataGroup:2022pth} for comparison.
We do not separately show any of the individual results
\cite{Gupta:2018qil,Walker-Loud:2019cif,Harris:2019bih,Liang:2018pis,Chang:2018uxx} 
that have entered the FLAG averages.
We note that our results are very competitive with regard to overall accuracy,
and more precise than the FLAG average in each case where an average exists. In
particular in the case of $g_T^{u-d}$, our result is more precise than any
of the competing determinations, while being entirely compatible with all of
them.

We note that the use of the NLO summation method, which we consider to be
superior in suppressing excited states both to the plain summation method and
to two-state fits to the ratios, is a crucial ingredient in the accuracy
achieved here.

Looking forward, we expect to obtain a similar improvement in precision for the
isoscalar charges of the nucleon, for which we possess all required ingredients,
given our ability to compute quark-disconnected loops to high precision
\cite{Agadjanov:2023efe,Djukanovic:2023jag}.

\section*{Acknowledgments}
This research is partly supported by the Deutsche Forschungsgemeinschaft (DFG,
German Research Foundation) through project HI 2048/1-2 (project No.~399400745),
and in the Cluster of Excellence \emph{Precision Physics, Fundamental
Interactions and Structure of Matter} (PRISMA+ EXC~2118/1) funded by the DFG
within the German Excellence strategy (Project~ID~39083149).
The authors gratefully acknowledge the Gauss Centre for Supercomputing e.V.
(\url{www.gauss-centre.eu}) for funding this project by providing computing
time on the GCS Supercomputers JUQUEENi~\cite{juqueen} and JUWELS~\cite{JUWELS}
at Jülich Supercomputing Centre (JSC) through projects CMHZ21, CHMZ23, CHMZ36,
NUCSTRUCLFL and GCSNucl2pt.
Additional calculations were carried out on the local HPC clusters ``Clover''
and ``HIMster2'' at the Helmholtz Institute Mainz, and ``Mogon 2'' at Johannes
Gutenberg University Mainz (\url{https://hpc.uni-mainz.de}), which is a member
of the AHRP (Alliance for High Performance Computing in Rhineland Palatinate,
\url{https://www.ahrp.info}), the Gauss Alliance e.V., and the NHR Alliance
(Nationales Hochleistungsrechnen, \url{https://www.nhr-verein.de}).
The QDP++ library \cite{Edwards:2004sx} and the deflated SAP+GCR solver from
the openQCD package \cite{openQCD} have been used in our simulation code, 
while the contractions have been explicitly checked using \cite{Djukanovic:2016spv}. 
We thank our colleagues in the CLS initiative for the sharing of gauge field
configurations.

\bibliographystyle{h-physrev}
\bibliography{refs}

\clearpage

\end{document}